\documentclass[12pt,preprint]{aastex}
\usepackage{graphicx,epsfig,wrapfig,paralist}
\usepackage{lscape}
\usepackage{rotating}
\newcommand{\bq}{\begin{equation}}
\newcommand{\eq}{\end{equation}}

\def\gtsim{\lower.5ex\hbox{$\buildrel > \over\sim$}}
\def\ltsim{\lower.5ex\hbox{$\buildrel < \over\sim$}}

\def\apjl{ApJL}
\def\apj{ApJ}
\def\apjs{ApJS}
\def\mnras{MNRAS}
\def\araa{ARAA}
\def\aj{AJ}
\def\aap{A\&A}
\def\aaps{A\&A Suppl.}
\def\nat{Nature}

\shorttitle{Rotating progenitors of PISNe}
\shortauthors{Chatzopoulos,Robinson,Wheeler}
\begin{document}
\title
{EFFECTS OF ROTATIONALLY-INDUCED MIXING IN COMPACT BINARY SYSTEMS WITH LOW-MASS SECONDARIES AND IN SINGLE SOLAR-TYPE STARS}
\author{E. Chatzopoulos\altaffilmark{1}, Edward L. Robinson\altaffilmark{1}, \& J. Craig Wheeler\altaffilmark{1}}
\email{manolis@astro.as.utexas.edu}
\altaffiltext{1}{Department of Astronomy, University of Texas at Austin, Austin, TX, USA.}

\begin{abstract}
 
Many population synthesis and stellar evolution studies have 
addressed the evolution of close binary systems in which the 
primary is a compact remnant and the secondary is filling its 
Roche lobe, thus triggering mass transfer. Although tidal 
locking is expected in such systems, most studies have neglected
the rotationally-induced mixing that may occur. Here we study
the possible effects of mixing in the mass-losing stars 
for a range in secondary star masses and metallicities. We find
that tidal locking can induce rotational mixing
prior to contact and thus affect the evolution of the secondary
star if the effects of the Spruit-Tayler dynamo are included both for
angular momentum and chemical transport. 
Once contact is made, the effect of mass transfer tends to be
more rapid than the evolutionary time scale, so the effects of
mixing are no longer directly important, but the mass transfer
strips matter to inner layers that may have been affected by
the mixing. These effects are enhanced for secondaries of 1-1.2~$M_{\odot}$ 
and for lower metallicities. We discuss the possible implications for
the paucity of carbon in the secondaries of the cataclysmic
variable SS Cyg and the black hole candidate XTE~J1118+480 and 
for the progenitor evolution of Type Ia supernovae.
We also address the issue of the origin of blue straggler stars in
globular and open clusters. We find that for models that include rotation
consistent with that observed for some blue straggler stars, evolution
is chemically homogeneous. This leads to tracks in the HR diagram that 
are brighter and bluer than the non-rotating main-sequence turn-off point. 
Rotational mixing could thus be one of the factors that contribute to the 
formation of blue stragglers.
\end{abstract}


\vskip 0.57 in

\section{INTRODUCTION}\label{intro}

Cataclysmic variables (CVs) have been recognized as binary systems with white dwarf primaries
and studied for decades (Warner 2003). A closely related field, 
the study of high and low-mass X-ray binaries (HMXBs and LMXBs respectively) containing neutron stars 
and black holes in close binary systems is somewhat younger, but also 
mature (Tanaka \& Lewin 1995). The general scenario invoked for the 
low-mass systems includes evolution through a common envelope phase then 
subsequent shrinking of the orbit through magnetic braking or emission 
of gravitational radiation. When the orbit period becomes short enough, hours to days, 
even unevolved main sequence companions can 
fill their Roche lobes and transfer mass to the compact primaries. 

This scenario has been studied quantitatively with stellar evolution and 
population synthesis codes. The close binary orbits imply that tidal forces 
should lock the secondary stars into rotational synchronism with the orbital 
period and the implied rotation can be quite rapid. This synchronization mechanism
is efficient for secondaries with convective envelopes. This rapid 
rotation may, in turn, induce rotational mixing. Rotational mixing in this 
context has recently been studied by De Mink et al. (2009) and Brott et al. 
(2011a) in high-mass systems, but most studies of compact binaries with 
low-mass secondaries have neglected this effect (e.g. Willems \& Kolb 2004; 
Han \& Podsiadlowski 2004; Belczynski et al. 2008; Lipunov et al. 
2009). Ergma \& Antipova (1999) invoke ``deep mixing" to account 
for the apparent hydrogen-deficiency of SAX~J1808.4-3658, but give no
physical model. They note that MXB~1916-05 ($P_{orb}$ = 50 min) and 
MX~1820-30 ($P_{orb}$ = 11.4 min) seem to require a Roche-lobe filling 
helium-rich secondary in order to explain its short orbital period 
(Paczynski \& Sienkewicz, 1981). 

Carbon deficiency may be another symptom of rotationally-induced mixing.
Examples of carbon deficient secondaries include 
the  black hole candidate XTE~J1118+480, which shows a distinct paucity of carbon
since there is no C~IV emission detected (Haswell et al. 2002; Gelino et al. 2006)
and the black hole X-ray binary A0620-00 in which 
the C~IV doublet is found to be anomalously weak compared to the other lines, consistent with the low 
carbon abundance deduced from NIR spectra of the secondary star in the system (Froning et al. 2007, 2011).
Low-mass carbon deficient secondaries have also been detected in many CV systems.
Harrison et al. (2004) concluded that the K dwarf secondary
in the CV SS Cyg was deficient in carbon. Bitner, Robinson \& Behr (2007) provided
estimates for the masses of the secondary star and the white dwarf (WD) in SS Cygni
($M_{WD} =$~0.81~$M_{\odot}$, $M_{s} =$~0.55~$M_{\odot}$) and presented a discussion on the 
unusual evolutionary characteristics of the secondary and Sion et al. (2010) presented
an extended analysis of its carbon poor UV spectrum.
Similar behavior has been observed for the CV AE Aqr (Jameson, King \& Sherrington 1980;
Eracleous et al. 1994). 
Harisson, Osborne \& Howell (2005) presented more cases of CVs with weak CO absorption lines
in their spectra, implying low C abundance, and suggested that material that has been
processed through the CNO cycle is finding its way into the photospheres of secondary stars. More
systems with close to zero C~IV emission were discussed in G{\"a}nsicke et al. (2003).
In a potentially related context,
Ibano et al. (2012) discuss the carbon deficiency in the mass-accreting 
primaries of Algol systems and attribute it to a carbon deficiency due 
to CNO burning on the lower-mass, mass-losing secondary. They do not 
suggest any specific mechanism for the carbon deficiency of the secondary.

Extensive discussions of the reasons for carbon depletion in the
secondaries of CVs are presented in 
Harrison, Osborne, \& Howell (2005) and Schenker et al. (2002). 
As argued in those works, it is possible to greatly deplete carbon by processing it in the massive stars before Roche lobe 
contact and then stripping off just enough of the outer layers during thermal timescale mass transfer (TTMT) to reveal layers depleted in carbon.
The depletion is estimated to be up to a factor of 100 in these models. This may actually happen in some CVs like AE~Aqr 
since there are other reasons for suspecting the secondary in that system is evolved. 
Nonetheless, there are two problems invoking the mass-stripping mechanism as the only cause of carbon depletion in some CVs. 
First, it could be that a high fraction of CVs show the carbon anomaly.  If so, current population synthesis 
models cannot account for such a high fraction of higher-mass progenitors (Harrison, Osborne \& Howell 2005). Second, the Schenker et al. (2002) mechanism 
requires fine tuning.  The initial stellar masses, orbital periods, and evolutionary stages need to be exactly right 
at first Roche lobe contact to obtain results that agree with observations.

These issues motivated us to explore the possibility that carbon is depleted by rotationally-induced
mixing that processes surface material through the core, where it is subjected to partial CN burning. 
Rotational mixing allows the secondaries to start with lower mass and does not seem to require fine tuning.
Sufficient
mixing can lead to quasi-homogeneous evolution and perhaps to low-mass
helium star secondary stars. This possibility is of general interest, but
may be especially interesting in the context of the progenitors of Type Ia supernovae. 
Rotationally induced mixing may also be relevant in cases of single stars such as
the ``blue straggler" stars (BSSs) in open and globular
clusters. Many BSSs are observed to rotate rapidly and to have depleted C/O abundances. 
Some BSSs may be slow rotators, but as we describe in \S3.3, some are observed to be
rapid rotators. That seems sufficient to mention the possibility of rotationally-induced
mixing without knowing the origin of the rotation, nor specifically invoking binary tidal locking
as the origin of the rotation. We thus also explore the possibility of rotationally-induced mixing in BSSs.

As remarked above, rotationally-induced mixing in tidally-locked massive 
stars has been investigated in some contexts (de Mink et al. 2009; Brott
et al. 2011a). Rotationally-induced mixing in massive stars has been 
investigated in general (Maeder 1987; Maeder \& Meynet 2011; Eskstrom et al.
2008, 2011; Brott et al. 2011b; Yoon, Dierks \& Langer 2012), in the context 
of the progenitors of gamma-ray bursts (Heger, Woosley \& Spruit 2005; 
Yoon \& Langer 2005), and in the context of very massive stars susceptible
to the electron/positron pair instability by Chatzopoulos \& Wheeler (2012)
and by Yoon, Dierks \& Langer (2012). Because massive stars tend to
be more radiation pressure dominated, they are closer to neutral dynamical
stability, and hence, all else being equal, easier to mix. Very low mass
stars are thought to be fully convective, and hence to mix spontaneously,
but are also so long lived that they do not evolve in the brief history of
the Universe. The case that interests us here are stars of modest mass,
of order the solar mass. These stars have outer convective, mixing, 
envelopes, but inner radiative cores that may resist rotationally-induced
mixing. The issue requires quantitative investigation and that is the
subject of this paper.

In \S2 we describe the stellar evolution models. We present our results
in \S3 and discuss the implications and conclusions in \S4. An
appendix presents a calibration of rotating, magnetic MESA models with
work in the literature.

\section{MODELS}\label{mods} 

We have used the Modules for Stellar Experiments in Astrophysics (MESA version 3647; Paxton et al. 2011) code to calculate the evolution
of a grid of low-mass secondaries (0.8~$M_{\odot}$, 1~$M_{\odot}$, 1.2~$M_{\odot}$, 1.5~$M_{\odot}$ and 1.8~$M_{\odot}$) 
for three different metallicities:
$Z =$~$Z_{\odot}$, $Z =$~0.1~$Z_{\odot}$ and $Z =$~0.01~$Z_{\odot}$. Sub-solar metallicity is encountered in the Large and Small
Magellanic Clouds (LMC and SMC respectively) and in globular clusters (GCs) 
and is relevant to CVs and binary systems found there. Models with sub-solar metallicity might also be relevant to
Type Ia SN progenitors in other environments. All models were run for two different
degrees of ZAMS rotation: 0 and 30\% of the critical Keplerian rotation $\Omega_{crit} =$~$(g(1-\Gamma)/R)^{1/2}$ where
$g =$~$GM/R^{2}$ is the gravitational acceleration at the surface of the star, $G$ is the gravitational constant, $M$ is the
mass, $R$ is the radius of the star and $\Gamma=L/L_{Ed}$ is the Eddington factor where $L$ and $L_{Ed}$ are the total radiated
luminosity and the Eddington luminosity, respectively. Initial (ZAMS) rigid body rotation was assumed in all cases.

The reason we did not consider
higher rotational velocities is because for a variety of mass ratios $q =$~$M_{2}/M_{WD}$ for the binary system,
the maximum allowable rotation is about 33-38\% the critical value, where $M_{2}$ is the mass of the secondary star and $M_{WD}$ 
is the mass of the white dwarf. This comes from combining Kepler's third law expressed in terms of angular velocity:
\begin{equation}
(\frac{\Omega_{Kep}}{\Omega_{crit}})^{2} = (R/a)^{3} (\frac{1+q}{q}),
\end{equation}
with the the Paczynski (1971) expression for the Roche lobe radius:
\begin{equation}
(R_{L}/a) = 0.38 + 0.20 \log(q),
\end{equation}
where $\Omega_{Kep}$ is the Keplerian orbital frequency, $R_{L}$ is the Roche lobe radius, $R$ is the radius of the secondary star
and $a$ is the orbital separation. Setting $R=R_{L}$ in Equation 2 to account for the fact that the maximum rotation
prior to mass transfer occurs when the secondary fills its Roche lobe,
Equations 1 and 2 yield:
\begin{equation}
\frac{\Omega_{max}}{\Omega_{crit}} = [0.38+0.2 \log(q)]^{3/2}(\frac{1+q}{q})^{1/2}.
\end{equation}
The dependence on $q$ is weak. For the range 1~$<q<$~3 we get 0.33~$<\Omega_{max}/\Omega_{crit}<$~0.38. 
We note that there are binary systems in which the secondaries have smaller rotational periods than
the orbital period. Recent examples include some {\it Kepler} discoveries of binaries consisting 
of A-type star secondaries and white dwarf primaries 
(van Kerkwijk et al. 2010; Breton et al. 2011; Carter et al. 2011; Bloemen et al. 2012). We do not consider 
this possiblity here, but the issue of rotational mixing in such systems is clearly of interest.

MESA was run with the Schwarzschild criterion for convection with the fiducial value for the mixing length parameter, 
$\alpha_{MLT}=$~1.5. Wind-driven mass loss was calculated using the prescriptions of
Glebbeek et al. (2009) and de Jager, Nieuwenhuijzen \& van der Hucht (1988) as implemented in the code.
MESA employs a combination of equations of state (EOS), but for the regime of densities and temperatures encountered
by low-mass stars the OPAL EOS tables of Rogers \& Nayfonov (2002) are used. For higher densities and temperatures the EOS transitions
to HELM EOS (Timmes \& Swesty 2000). For the treatment of nuclear processes with MESA we employ the ``approx21" network (Timmes 1999), 
which covers all major stellar nuclear reaction rates. The effects of angular momentum and chemical transport via rotation and magnetic fields are 
treated based on the one-dimensional approximations of Spruit (1999, 2002), Heger, Langer \& Woosley (2000) and
Heger, Woosley \& Spruit (2005). Those include different types of mixing due to rotation such as dynamical shear instability,
secular shear instability, Solberg-Hoiland instability, Eddington-Sweet circulation (meridional circulation), Goldreich-Schubert-Fricke
instability as well as the Spruit-Tayler (ST) dynamo. For the efficiency of rotationally-induced mixing, $f_{c}$, we adopt the value 0.0228 based
on the findings of Brott et al. (2009) who calibrated this parameter using recent observations of massive rotating early B type stars in the LMC
and the SMC. 
A value of $f_{c}$ of 0.046 was proposed by Pinsonneault et al. (1989) in order to explain solar lithium abundances 
while theoretical predictions by Chaboyer \& Zahn (1992) provided
a value of 0.033, which was adopted in calculations of massive rotating models done by Heger et al. (2000). We note that most of those
values are calibrated by observations of rapidly rotating massive nearby stars. There are currently
no observations of rapidly rotating ($>$~20-30\% the critical velocity) solar-type stars that can yield a value of $f_{c}$ directly relevant to them.
Rotationally-induced mass loss is also calculated following
Heger, Langer \& Woosley (2000) and is equal to $\dot{M}_{rot}=\dot{M}_{no-rot}/(1-\Omega/\Omega_{crit})^{0.43}$ where $\dot{M}_{no-rot}$ is
the mass loss rate in the case of zero rotation, due to the effect of radiatively driven winds. In order to take into
account the inhibiting effect of a gradient in the mean molecular weight in the efficiency of mixing triggered by rotation we 
adopted the value $f_{\mu} =$~0.1 which is the same as the one used by Yoon et al. (2005) who calibrated their models 
including the effects of ST.

In order to validate the results found with the new MESA stellar evolution code that includes the effects of
rotation and magnetic fields against older well-established codes such as the Geneva stellar evolution code (Eggenberger et al. 2008), 
the Yoon \& Langer (2005) code and the modified KEPLER code introduced by Heger, Woosley \& Spruit (2005) we ran some of the same models presented
in those works with MESA. We present an extensive comparison of the results in Appendix A. There, we also compare our results with 
observations of nearby massive rotating stars provided by the VLT-FLAMES survey (Hunter et al. 2008, 2009).

In our suite of models, we find that the mixing is dominated by the meridional circulation and the ST mechanism, 
with the Goldreich-Schubert-Fricke instability having a minor role. 
The contribution of the ST mechanism tends to
be comparable to, or even slightly dominant over, the meridional circulation at mass fraction greater than 0.8 - 0.9, and to
dominate in a shell covering the mass fraction between 0.2 and 0.4. 
While the basic physics of the ST mechanism
is solid, the employment of this complex, multi-dimensional process in a spherical model may be especially suspect. 
We checked the role of this process by omitting the Spruit-Tayler mechanism in a trial calculation of the model with 
1.0~$M_{\odot}$, and $Z =$~$Z_{\odot}$ rotating at 30\% of the critical velocity (see Appendix A, Figure A7). 
Omission of the Spruit-Taylor mechanism changed
our results considerably. We find that meridional mixing alone in low mass stars cannot induce significant surface abundance 
changes that are adequate to explain the observed features of carbon-deficient binaries. We return to this point in
the conclusions.

We define a timescale, $t_{RL}$, as the time for the secondary star to evolve from non-contact to Roche lobe overflow (RLOF). 
This time-scale is ill-constrained due to uncertainties involved
in the mechanism of angular momentum loss (common envelope evolution, magnetic braking) 
and can vary over a large range. We will
discuss those uncertainties in more detail in \S 3.2.
As a fiducial value for the start of RLOF, chosen in order to calibrate our results for different initial masses and present
surface abundance ratios at this epoch in Table 1,
we choose $t_{RL} \simeq$~$\tau_{MS}$, where $\tau_{MS}$ is the main-sequence life time of the secondary.
Once our models reach RLOF, a constant mass loss rate typical for CVs can be employed, in the range 2-6~$\times 10^{-10}$~$M_{\odot}$~yr$^{-1}$.
This rate of mass loss reduces the secondary to a mass typically around 0.6~$M_{\odot}$, 
characteristic of white dwarf - red dwarf binary systems such as SS Cygni, within $\sim 3 \times 10^{9}$~years. 
This mass loss rate is scaled up by four orders of magnitude if mass loss
occurs on a thermal timescale, a scenario that we discuss in \S 4.
We emphasize that we consider the evolution of the secondary stars as single stars, and we do not
model the evolution of the binary system as a whole.
In the case of close binary systems, the secondary stars will experience 
tidal forces that will change the shape of the star. 
Rotation will have the additional effect of inducing oblateness. Those effects are not considered in our one-dimensional stellar
evolution calculations.

\section{RESULTS}\label{results}

\subsection{{\it Evolution of Surface Composition.}}

 In order to illustrate the general effects of rapid rotation on the evolution of low-mass stars, we present the evolution of 
$\Omega/\Omega_{crit}$ with time in the left panel and the evolution in the H-R diagram in the right panel of Figure 1 for
the 1~$M_{\odot}$, $Z =$~$Z_{\odot}$ model. MESA includes solar mixtures on several scales as presented by 
Asplund et al. (2005).
We see that $\Omega/\Omega_{crit}$ slowly increases 
with time during the MS. This is the result
of increasing radius and luminosity and can be shown by considering angular 
momentum conservation and the definition of $\Omega_{crit}$ as presented
in \S 2. Even though $\Omega/\Omega_{crit}$ increase, it remains under the upper limits 
of 0.33-0.38.
As is the case with more massive rotating models, the low-mass rotating stars in the HR diagram lie everywhere bluer than the evolutionary
tracks of the non-rotating stars.

 Next, we examine the effect of rapid rotation on the surface abundances of key elements such as $^{1}$H, $^{4}$He, $^{12}$C and $^{14}$N. 
Sufficient rotational mixing is expected to lead to chemically homogeneous evolution, as observed in more massive stars 
(Heger, Langer \& Woosley 2000; de Mink et al. 2009; Chatzopoulos \& Wheeler 2012; Yoon, Dierks \& Langer 2012). 
Mixing of material from the core up to the surface leads to different
surface composition than for the non-rotating case. In general, models of rapidly rotating stars are found to be rich in 
$^{14}$N and $^{4}$He while heavily depleted in $^{1}$H and less so in $^{12}$C. Spectroscopy of rapidly rotating massive stars seems 
to indicate, however, that the efficiency of rotational mixing is still controversial (Hunter et al. 2008; Brott et al. 2011b) since
some discrepancies are found with the predictions of some rotating models in the evolution of surface abundances.

 Figures 2 through 5 show the evolution of the surface mass fractions of 
$^{1}$H, $^{4}$He, $^{12}$C and $^{14}$N for the grid of models we investigate
in this work. In each figure, the upper left panel shows the results for $Z =$~$Z_{\odot}$, the upper right panel for 
$Z =$~0.1~$Z_{\odot}$ and the lower left panel for $Z =$~0.01~$Z_{\odot}$. In all cases solid lines correspond to the 
$\Omega/\Omega_{crit} =$~0 models and dashed lines to $\Omega/\Omega_{crit} =$~0.3 models. Also, black color denotes  
$M_{2} =$~0.8~$M_{\odot}$, red 1~$M_{\odot}$, blue 1.2~$M_{\odot}$, green 1.5~$M_{\odot}$ and orange 1.8~$M_{\odot}$.
Figure 6 shows the evolution of the stellar radius with time for all the models and also provides information on the MS life-time of each model
before they enter the giant phase. The steep vertical lines for the non-rotating models in Figures 2 - 5 represent the onset of 
evolution to the red giant branch, as shown in Figure 6. This transition is muted or modified for the rotating models. Figures
2 - 6 show that rapidly rotating stars generally live longer than those of zero rotation. All evolutionary tracks have been computed
almost up to core helium ignition.

 Figure 2 shows the evolution of the surface abundance of hydrogen, $X_{H,s}$, for all the models. The depletion of hydrogen
is stronger for intermediate masses and lower metallicities. This is a natural consequence of the fact that somewhat higher mass stars have hotter
interiors so that nuclear burning via the pp-chain and CNO cycle processes is more efficient, but as the mass gets even higher, the star
lives on the main-sequence for a smaller time that is not sufficient for mixing before it comes in contact or before turning into a
giant. The 1~$M_{\odot}$, 
$Z =$~0.01~$Z_{\odot}$, $\Omega/\Omega_{crit} =$~0.3 model reaches $X_{H,s} <$~0.62 by the end of the MS. 
 In Figure 3, the same model is seen to be helium enriched on the surface, compared to the non-rotating
case.
The outcome of enhanced CNO processing in rotating stars is best seen in Figure 4 where depletion of carbon at the surface
occurs towards high rotation, intermediate masses and lower metallicity and in Figure 5 where the resulting $^{14}$N enrichment is seen in the same
regions of the parameter space. The results for all models are given in Table 1 where the ratio of surface mass fraction for all those elements on the ZAMS
compared to that on the end of the MS is given.

The effects of rotation on the composition structure of low mass secondaries are 
also illustrated in Figure 7.
We observe that, even though homogenization is weaker 
in low mass stars than it is in more massive stars,
the partial mixing that occurs is strong enough to drive abundance changes throughout the interior of the star. 
This also results in the cores of the rotating stars being
somewhat more massive that those of non-rotating ones and in
hydrogen depletion at the center being more pronounced, especially for lower metallicities.

In order to investigate the effects of RLOF mass loss on the evolution of the surface abundances of our models, we have run the 
1~$M_{\odot}$, $Z =$~$Z_{\odot}$, $\Omega/\Omega_{crit} =$~0 and the  
1~$M_{\odot}$, $Z =$~$Z_{\odot}$, $\Omega/\Omega_{crit} =$~0.3
models up to $t_{RL}$~$\sim$~$5 \times 10^{9}$ years without RLOF mass loss 
and then turned on a constant mass loss rate of $5 \times 10^{-10}$~$M_{\odot}$~yr$^{-1}$, within the range expected for CVs. 
This mass loss is maintained until
the total mass of the star is reduced to $\sim$~0.55~$M_{\odot}$, requiring an interval of $\sim 9 \times 10^{8}$~years. 
The evolution of the surface mass fraction of carbon, $X_{C,s}$, for
these models is shown as the solid (non-rotating) and dashed (rotating) light blue curves in the upper left panel of Figure 4.  
In the case of the non-rotating mass-losing model, $X_{C,s}$ does not change much
because carbon depletion is strong only in the inner ~0.3~$M_{\odot}$ of the star and mass-stripping does not reach this depth. Also,
since we are left with a 0.55~$M_{\odot}$ star its MS life-time is significantly prolonged leaving $X_{C,s}$ at the same value
for more than a Hubble time.
For the rotating, mass-losing model we see that
RLOF mass loss has 
only a small effect, a decrease of about a factor of 2, on the further evolution
of the surface carbon abundance. However, 
the comparison between the mass-losing non-rotating and rotating models indicates that rotationally-induced 
mixing leading to abundance changes throughout the whole stellar interior remains the main reason for $X_{C,s}$ depletion even
when mass loss is accounted for, at least for the fiducial example presented here. It is the mixing that took place the first 
$\sim$~$5 \times 10^{9}$ years that lead to carbon depletion in the surface as well as in some inner layers in different degrees with mass loss
being a second order effect because in the absence of such mixing, as can be seen for the non-rotating mass-losing model, mass loss
alone did not lead to significant changes. For both mass-losing cases we also note that
once the stellar mass becomes small ($< 0.8$~$M_{\odot}$), carbon depletion is diminished because
the core temperature becomes sufficiently low that the fusion of C to N can no longer take place. In addition,
rotationally-induced mixing becomes less efficient.
In reality, the process of RLOF mass loss is, of course, more complicated and the mass loss rate changes with time such that it starts
with initially very high mass loss rate then declines to values characteristic for CVs.
We argue that the most interesting phase of chemically quasi-homogeneous evolution in rapidly rotating secondaries
is that before the star loses significant amounts of mass via RLOF. The extent of RLOF mass loss itself 
has an impact only on the magnitude of these effects in the sense that deeper regions of the star are probed with somewhat different
(but not significantly so) composition.

\subsection{{\it Constraints by Loss of Angular Momentum.}}

 The effects of rotationally-induced mixing on the surface abundances of low mass secondaries that we presented is \S 3.1 are only valid
under the assumption that the secondary star rotates at relatively high velocities ($\sim$~30\% the critical value) for a sufficient time,
$t_{RL}$, before loss of angular momentum causes the secondary to fill its Roche lobe and begin transfering mass to the white dwarf. 
To investigate
the constraints of angular momentum loss on our results we calculate relevant time-scales. The time scales
we wish to compare are the characteristic time-scales for synchronization, for sufficient mixing that leads to considerable surface abundance 
changes, and for angular momentum loss.

First of all, we demand that the system be synchronized so that the orbital period is equal to the rotational period of the 
secondary. The synchronization time-scale for a binary system based on the influence on tides
is given by the following expression (Zahn 1977, 1989):
\begin{equation}
\tau_{sync} = f_{turb} \frac{1}{q^{2}} (\frac{a}{R_{2}})^{6}\mbox{~yr},
\end{equation}
where $f_{turb}$ is a constant that depends on the structure of the secondary and is typically of the order of unity for stars with convective
envelopes. 
For a combination of masses and for $a/R_{2}\leq$~20 the synchronization time-scales are less than $\sim 10^{8}$~years, indicating
that most of the systems under consideration synchronize rapidly compared to their MS life spans. 
This time-scale can be even shorter if radiative dissipation due to oscillation damping near the stellar surface is considered (Zahn 1975).
For the mixing time-scale, we choose a lower limit of $t_{RL} = 3 \times 10^{9}$~years and an upper limit of 
$t_{RL} = 7 \times 10^{9}$~years based on the maximum
carbon depletion that we find for the representative models of 1~$M_{\odot}$ and 1.2~$M_{\odot}$ for $Z =$~0.1~$Z_{\odot}$ and 
$Z =$~0.01~$Z_{\odot}$ ($X_{C,s}$ declines by about an order of magnitude for these models in the chosen time range). 

The angular momentum loss time-scale is given by $\tau_{AML}=J/\dot{J}$ where $J$ is the total orbital angular momentum and $\dot{J}$ is
the rate of change of the orbital angular momentum. 
We can write
$J = M_{\odot}^{5/3} G^{2/3} m_{wd} m_{2} m^{-1/3} \Omega_{orb}^{-1/3}$ with $M_{\odot}$ the solar mass, $G$ the gravitational constant,
$m_{wd}$, $m_{2}$ and $m$ the masses of the white dwarf, the secondary star and the sum of the two in units of the solar mass, respectively 
(capital letters denote those masses in cgs units),
and $\Omega_{orb} = (G M_{\odot} m/a^{3})^{1/2}$ is the orbital frequency. The prescription for $\dot{J}$ depends
on the angular momentum loss mechanism. 

There are two basic models for angular momentum loss in binary systems: angular momentum loss via
gravitational radiation (for systems with shorter orbital periods) and via magnetic braking 
(for systems with larger orbital separation). 
For angular momentum loss by gravitational radiation Paczynski (1967) gives for $J/\dot{J}$:
\begin{equation}
\tau_{GR} = \frac{1.25 \times 10^{9} r^{4}}{m_{2} m_{wd} m} (\frac{a}{R_{2}})^{4}\mbox{~years},
\end{equation}
where $r=R_{2}/R_{\odot}$ with $R_{2}$ the radius of the secondary star in cm. 

Angular momentum loss by magnetic braking was introduced by Verbunt \& Zwaan (1981) and is based on the principle that angular momentum 
is removed from the secondary via magnetically coupled stellar winds. Then, from the orbital tidal synchronization and
using their calculated $\dot{J}$ we estimate $\tau_{AML}=\tau_{MB}$ for
this process to be:
\begin{equation}
\tau_{MB} = \frac{3.33 \times 10^{6} f^{2} m_{wd} m_{2} r}{k^{2} m^{3}} (\frac{a}{R_{2}})^{5}\mbox{~years},
\end{equation}
where $f$ is a constant of order of unity and $k=(0.1)^{1/2}$.
Rappaport et al. (1983) offered an updated prescription for angular momentum loss with magnetic braking accounting for the structure
of the radiative core and the convective envelope of the secondary. Using their method we calculate:
\begin{equation}
\tau_{MB} = \frac{4.38 \times 10^{6} m_{wd} m_{2} r^{5-\gamma}}{m^{3}} (\frac{a}{R_{2}})^{5}\mbox{~years},
\end{equation}
where $\gamma$ is a dimensionless parameter that ranges between 0-4. For $\gamma =$~4 an expression close to that of Verbunt \& Zwaan (1981)
is recovered. 
To better constrain the magnetic braking mechanism, Sills et al. (2000) obtained rotational data from young open clusters and calibrated
the theoretical predictions. They found that the mechanism is less efficient than previously thought. Using Equation 8 for $\dot{J}$ from their work, 
we derive the corresponding angular momentum loss time-scale:
\begin{eqnarray}
\tau_{MB} = \cases{\displaystyle \frac{2.88 \times 10^{6} m_{wd} m_{2} r^{9/2}}{m^{3/2}}(\frac{a}{R_{2}})^{5}\mbox{~yr}, & $\Omega \leq \Omega_{X}$, \cr \cr
\displaystyle \frac{1.13 m_{wd} m_{2} r^{3/2}}{m^{1/2} \Omega_{X}^{2}}(\frac{a}{R_{2}})^{2}\mbox{~yr}, & $\Omega > \Omega_{X}$,}
\end{eqnarray}
where $\Omega_{X} =$~$\Omega_{X,\odot} \tau_{conv,\odot}/\tau_{conv,2}$ is the critical frequency
at which the magnetic field apparently saturates (Krishnamurthi et al. 1997), $\tau_{conv,\odot}$ and $\tau_{conv,2}$ are the convective
overturn time-scales for the Sun and for the secondary star respectively. Sills et al. (2000) find that a value of 
$\Omega_{X} \simeq$~10~$\Omega_{\odot}$ is needed to fit the data, and we adopt the same value in this work. 
Ivanova \& Taam (2003) presented yet another refinement for the angular momentum loss based on the relation between X-ray variability and rotation
rate in X-ray binaries. Their relation also has the form of a broken power law and also depends on the magnetic field
saturation frequency. Their prescription implies a time-scale for angular momentum loss given by the following formula:
\begin{eqnarray}
\tau_{MB} = 
\cases{\displaystyle \frac{3.16 \times 10^{6} m_{wd} m_{2} r}{t^{0.5} m^{2}} (\frac{a}{R_{2}})^{5}\mbox{~yr}, & $\Omega \leq \Omega_{X}$, \cr \cr
\displaystyle \frac{11.35 m_{wd} m_{2}}{r^{1.55} t^{0.5} m^{1.15} \Omega_{X}^{1.7}} (\frac{a}{R_{2}})^{2.5}\mbox{~yr},
& $\Omega > \Omega_{X}$,}
\end{eqnarray}
where $t=T_{2}/T_{\odot}$, with $T_{2}$ the effective temperature of the secondary and $T_{\odot}$ the
effective temperature of the Sun. In order to calculate the numerical coefficients we have used $\Omega_{\odot} =$~$2.9 \times 10^{-6}$~rad~s$^{-1}$.

The more recent predictions for angular momentum loss by magnetic braking imply a less efficient mechanism than the older predictions.
Yet another correction to those laws comes from the fact that
stars with a small or no convective mantle do not have a strong magnetic field and will therefore experience little or no magnetic braking
(Podsiadlowski et al. 2002; van der Sluys, Verbunt \& Pols 2005). In this case, a correction term $\exp(0.02/q_{conv}-1)$ can be defined, where $q_{conv}$
is the fractional mass of the convective envelope. Adopting this correction to the law of Sills et al. (2000) (Equation 8) makes the effective angular
momentum loss time-scale even larger. Large angular momentum loss timescales of the order of a Hubble time due to magnetic braking have also 
been discussed by Andronov et al. (2003) (see Figure 4 of their paper) as well as by Willems et al. (2007). 
In the context of these long magnetic braking time-scales, the period gap for CVs with orbital period between
2 and 3 hours is addressed by invoking two distinct populations. As discussed in Andronov et al. (2003), 
one population would comprise unevolved secondaries (those with orbital periods of 2 hours or less)
and the other evolved secondaries (with orbital periods of 3 hours or more). 

The large time scales for magnetic braking posited in the more recent papers
imply that it is possible for some close systems to be synchronized and maintain nearly a constant orbital period for
long enough that their secondaries rotationally mix, therefore yielding altered surface mass fractions for some elements.
To illustrate this idea, we plot all the angular momentum loss time-scales discussed above as a function of orbital period in 
Figures 8 and 9. In Figure 8 we have done so for a system with $M_{2} =$~1.2~$M_{\odot}$ and $M_{WD} =$~0.65~$M_{\odot}$,
which we suspect is similar to SS Cyg prior to its contact with its Roche lobe.
The inset labels the various angular momentum loss time-scales for gravitational radiation and magnetic braking. 
It can be seen that for systems with $P_{orb} \leq$~7-8~hrs there is sufficient time for the secondary to undergo significant rotationally
induced mixing before coming in contact with its Roche lobe, if we accept the recently updated, less efficient magnetic braking laws. 
In general, for a secondary to sufficiently mix before transfering mass to the white dwarf, a fiducial system has to be in the region between
the two dotted lines and towards smaller periods, so that the rotational velocity is higher and therefore the rotationally-induced mixing
stronger. In addition, the system has be below the curves for angular momentum loss as calculated by making use of the latest
prescriptions for angular momentum loss via magnetic braking in the same diagram. 
In Figure 9 we present the same plot, but in the particular case of the system V471~Tau 
($M_{2} =$~0.9~$M_{\odot}$, $M_{WD} =$~0.84~$M_{\odot}$, $P_{orb} =$~12.5~hrs; O'Brien et al. 2001) consisting of a white dwarf and a red dwarf
secondary star. A lower limit for $J/\dot{J}$ in this case was estimated by using the observed $P_{orb}/\dot{P}_{orb}$ coming from
the observed minus corrected ($O-C$) diagram. 
It can be seen that it is possible that
this system will mix in the future; however, at that large separation, it probably will not
exhibit a significant rotationally-induced mixing due to the low orbital frequency. 

We conclude that the type of binary system with low-mass secondaries that might undergo significant rotationally-induced mixing
that leads to enhanced nitrogen and helium and depleted hydrogen and carbon surface abundances is
one that evolves in a manner illustrated in Figure 10. Initially (panel A) the system consists of a higher mass primary ($M_{1}$) and
a lower mass secondary ($M_{2}$). Subsequently the system undergoes common envelope evolution and the primary dies, becoming a white dwarf, while
the secondary is still on the main-sequence (panel B). At this point the system is synchronized and the orbital separation is relatively
small ($P_{orb} \leq$~7-9~hrs). The system remains detached and the secondary rotates at high velocities 0.2~$\leq \Omega/\Omega_{crit} \leq$~0.33 
thus undergoing rotationally-induced mixing (panel C). Those fiducial orbital period and corresponding rotational angular velocity limits
are implied for significant surface carbon depletion, as shown in Figure 8.
The mixing significantly alters the surface element abundances by the
end of the MS, leading to the depletion of carbon and the enchancement of nitrogen in the surface. Finally (panel D), the system 
has lost sufficient angular momentum that the secondary comes in contact with its Roche lobe and starts to transfer mass to the
white dwarf. A long timescale between tidal locking and RLOF is consistent with current theories of relatively weak 
magnetic braking. At the onset of RLOF, unstable mass transfer could occur on the thermal time-scale of the secondary or stable
mass transfer could ensue for mass ratio, $ q < 1$ on long time-scales characteristic of many CVs, as we have assumed
for the model here. The mass transferred to the WD
has different abundances than in cases where the secondary does not rotate rapidly and thus does not mix enough,
or if the secondary fills its Roche lobe before mixing is extensive. 

\subsection{{\it Application in the Case of Blue Straggler Stars.}}

The inclusion of the effects of rotation in the evolution of solar-type stars may also provide insight to some cases of
blue straggler stars (BSSs) first discovered by Sandage (1953) in the GC M3. 
BSS are stars in open or globular clusters that appear to be on the MS but are more luminous and bluer than stars at the MS turn-off 
point in the cluster HR diagram. BSS have been found in many 
globular and open clusters in the Milky Way, notably the globular cluster
47~Tuc (Ferraro et al. 2006), the open cluster NGC~188 (Mathieu \& Geller 2009) and the globular cluster M4 (Lovisi et al. 2010).

Suggestions for the origin of BSSs include mass transfer from primordial binaries in low density GCs (MT-BSSs) or even
stellar collisions, particularly those that involve binaries (COL-BSSs) in high star density GCs 
(McCrea 1964; Hills \& Day 1976; Fusi Pecci et al. 1992; Ferraro et al. 2003; Davies et al. 2004). There is some evidence that both
processes could be active in some clusters (Ferraro et al. 2009). There are still open issues concerning the surface chemical abundances
of BSSs. In particular, depleted C/O abundances are associated with BSSs in some clusters, for example in 47~Tuc (Ferraro et al. 2006).
In addition, high equatorial velocity has been measured for many BSSs: the equatorial velocity in 47~Tuc is measured to be 
$>$~10~km~s$^{-1}$ (Ferraro et al. 2006), in NGC~188 many BSSs have velocities of up to 50~km~s$^{-1}$ (Mathieu \&
Geller 2009) and in M4 the BSSs equatorial velocities range from 10~km~s$^{-1}$ up to 150~km~s$^{-1}$ ($\Omega/\Omega_{crit}\sim $~0.02-0.35)
(Lovisi et al. 2010).

The depleted C/O abundances and relatively high rotation that are observed in some BSSs indicate that rotationally-induced mixing leading
to chemically homogeneous evolution might be relevant for some of those objects. The idea of prolonged strong mixing associated with BSSs
was first suggested by Wheeler (1979; see also Saio \& Wheeler 1980). With the availability of stellar evolution codes 
that take into account the effects of rotation, we can attempt to evaluate this hypothesis. 

We ran MESA with 1~$M_{\odot}$
models for solar ($Z =$~$Z_{\odot}$), 10\% solar ($Z =$~0.1~$Z_{\odot}$) and 1\% solar ($Z =$~0.01~$Z_{\odot}$) metallicity 
for four different rotation velocities in the range
of those measured for the BSSs in the clusters 47~Tuc, NGC~188 and M4: $v_{rot} =$~0~km~s$^{-1}$, 10~km~s$^{-1}$, 60~km~s$^{-1}$ and 
100~km~s$^{-1}$. 
The evolution of these models in the HR diagram is shown in the upper left ($Z =$~$Z_{\odot}$), the upper right ($Z =$~0.1~$Z_{\odot}$)
and the lower left ($Z =$~0.01~$Z_{\odot}$) panels of Figure 10. 
In all panels, solid curves represent $v_{rot} =$~0~km~s$^{-1}$, dashed curves $v_{rot} =$~10~km~s$^{-1}$, dashed-dotted
curves $v_{rot} =$~60~km~s$^{-1}$ and dotted curves $v_{rot} =$~100~km~s$^{-1}$. The ``BSS" labeled boxes indicate where BSS candidate stars
are expected to be located in the HR diagram of a cluster. It is evident that the evolutionary tracks of stars with equatorial velocities 
$>$~10~km~s$^{-1}$ are well within the BSS domain. 
Given the fact that there is a spread in masses, compositions and equatorial velocities associated with BSSs we expect that the BSS labeled
regions in Figure 11 in reality will be even more densly populated with evolutionary tracks than the sample of models we
present here. 
We conclude that quasi-homogeneous evolution due to rotationally-induced mixing may be an alternative, third channel to form BSSs.
The rotation to induce the mixing may have arisen in mass-transfer events, so all these mechanisms may be active. 

\section{DISCUSSION AND CONCLUSIONS}\label{disc}

G{\"a}nsicke et al.(2003) presented a variety of CVs with very large N$_{V}/$C$_{IV}$ flux ratios in far ultraviolet (FUV)
spectra. In the same study it is estimated that 10-15\% of close CVs that have gone through a phase of TTMT 
show this kind of surface abundance anomalies.
Similar cases of large N$_{V}/$C$_{IV}$ flux ratios have been observed in some black hole binary systems
like XTE~J1118+480 (Haswell et al. 2002; Gelino et al. 2006). 
Most low-mass close binaries are also synchronized and their secondaries
are rapidly rotating and therefore evolve quasi-homogeneously due to rotationally-induced mixing.

Motivated by these observations, we ran a grid of evolutionary models of low mass secondaries
for different degrees of metallicity and rotational velocity in order to study the effects of rotationally-induced mixing in these systems.
Vigorous rotationally-induced mixing led to a more chemically homogeneous evolution
than in non-rotating standard evolution and enabled CNO processing of material in a larger portion of the stellar interior. As a consequence,
the hydrogen and carbon surface abundances are reduced and the helium and nitrogen abundances enhanced. This
effect is found to be stronger for intermediate mass (1-1.2~$M_{\odot}$), increased rotation, and decreased metallicity for the secondary star. 
The magnitude of this effect is also bigger for longer time-scales after tidal locking and before encountering RLOF mass loss. 
Once RLOF mass loss starts, the surface
abundances do not change significantly and by the time RLOF ends the secondary may be left with low enough mass that the core temperature
is not high enough to burn the material that is mixed inwards, therefore preventing further homogenization and significant surface abundance changes. 
Since carbon depletion is more pronounced
for higher masses, it is possible that the secondaries of carbon depleted systems were once more massive, even up to 1.2-1.5~$M_{\odot}$, and then lost a 
significant amount of mass. This can be achieved via a phase of unstable thermal time-scale mass transfer (TTMST; Schenker et al. 2002;
Podsiadlowski, Han \& Rappaport 2003; G{\"a}nsicke et al. 2003). In this case, systems with secondaries with mass up to 2~$M_{\odot}$
may subsequently evolve into CVs with low mass secondaries and slow transfer rates.  If the secondary were a fast rotator, 
the mass transferred will have a high ratio of
$X_{N,s}/X_{C,s}$, in agreement with what is spectroscopically observed.
Therefore we conclude that rotationally-induced mixing could be one way to explain the carbon-depletion
features of some of the systems discussed above. 

Altering the abundance of the material accreted onto the white dwarf may affect the subsequent
evolution of the white dwarf. Systems containing a white dwarf and a low mass helium-rich 
companion are observed (Maxted et al. 2000; Mereghetti et al. 2011). Helium overabundance is 
spectroscopically observed in some recurrent novae, U Sco (Williams et al. 1981; Hanes 1985; 
Starrfield et al. 1988) and V394 (Sekiguchi et al. 1989) and some classical novae, Nova LMC 1990 
no. 2 (Sekiguchi et al. 1990; Shore et al. 1991) and V445 Puppis (Nova Puppis 2000; Ashok \& Banerjee 2003; 
Kato \& Hachisu 2003; Kato et al. 2008; Woudt et al. 2009; Goranski et al. 2010). In most considerations 
in the literature, the helium-rich secondaries are the cores of stars stripped of their red-giant envelopes 
by mass transfer (Iben \& Tutukov 1994; Yoon \& Langer (2003); Solheim \& Yungelson 2005; Ruiter et al. 
2009, 2011; Wang et al. 2009a,b; Wang \& Han 2009; Meng \& Yang 2010). Here, we suggest that 
helium-rich secondaries can arise through rotationally-induced mixing of main sequence stars. Some 
of our models exhibit enhanced $X_{He,s}/X_{H,s}$ therefore providing accreting material of a different 
mixture than has been extensively modeled to date (solar H/He or pure He).

As discussed in \S 2 and in Appendix A, we find interesting levels of rotationally-induced mixing only if
we include the effects of the Spruit-Tayler dynamo on the transport of both angular momentum and 
chemical abundances. We suspect that rotation, especially differential rotation, will induce magnetic 
effects and that the omission of magnetic effects is inappropriate. Whether the Spruit-Tayler mechanism as employed
in shellular models is the ``correct" or only magnetic effect is not so clear. Inclusion of the Spruit-Tayler mechanism
in rotating stellar evolution calculations is ``state-of-the-art," and we include it in order to capture some magnetic effects 
and to compare to other work in the literature that makes comparable assumptions (Appendix A). 
We emphasize that the inclusion
of chemical mixing due to magnetic fields as parametrized by Spruit (2002) is observationally motivated in the present work since it is necessary 
in order to explain the observed surface abundance changes in some low-mass secondary stars, members of close binaries that are rapid
rotators, in the context of stellar evolution with the inclusion of the effects of rotation.
The inclusion of magneto-rotational effects in stellar evolution is a topic that surely warrants 
more attention (Brown et al. 2011).

The inclusion of the effects of rotation in the evolution of some low mass secondaries may
thus have implications for the progenitors of Type Ia SNe.
Accretion of hydrogen onto a white dwarf will tend to generate double shell sources that 
are susceptible to thermonuclear instabilities, including nova explosions.  Even with high 
accretion rates that allow steady hydrogen shell burning, the subsequent helium shell burning 
is often found to be unstable (Iben \& Tutukov 1989; Cassisi et al. 1998; Kato \& Hachisu 1999) 
making it very difficult to construct satisfactory models that grow the white dwarf to near the 
Chandrasekhar mass, carbon ignition and thermonuclear explosion. At low accretion rates, 
the helium shell source in accreting white dwarf models may be thermally unstable even for 
non-degenerate cases (Cassisi et al. 1998; Langer et al. 2002). Accretion from a hydrogen-poor 
secondary may modify some of these constraints (Webbink et al. 1987; Truran et al. 1988; Livio \& Truran 1992). 
A higher helium abundance will tend to lead to more stable shell burning. This might alter
the conditions of shell burning and hence the constraints necessary for sucessful Type Ia explosions. 
Based on our models, the $X_{He,s}/X_{H,s}$ ratio would increase during the accretion process 
since deeper regions of the rotationally mixed stars become accessible. Further exploration of
this issue requires thermonuclear hydrodynamic simulations. We note that although our rotating models
linger longer on the main sequence, the luminosity does increase monotonically. Models
that are more helium enriched will thus also tend to be somewhat brighter than ZAMS models
of the same mass. All of our models with noticeable helium enrichment are brighter than
the upper limit set on a particular SN Ia progenitor system by Schaefer \& Pagnotta (2012).

Helium detonations on white dwarfs have also been proposed to be related to
the progenitors of the predicted class of subluminous .Ia SNe (Shen et al. 2010). 
Events such as SN~2002bj (Poznanski et al. 2010) and SN~2010X (Kasliwal et al. 2010)
show strong He lines in their spectra, but no sign of H. A potential channel to .Ia progenitors
could be one of a close synchronized binary system with a rapidly rotating secondary star that
undergoes chemically homogeneous evolution leading to surface He enrichment. Once RLOF mass loss sets in,
this He rich material from the secondary will accrete on the WD and, provided that the accretion rate
is appropriate for stable shell burning, it may set the initial conditions appropriate for a .Ia progenitor. 

Rotationally-induced mixing seems to play a role in the evolution of solar mass rotating single stars as well.
As an example we examined the case of BSSs found in many galactic clusters. We found that the inclusion of the effects of rotation
in the evolution of solar-type stars might lead some of those with equatorial rotational velocity greater than 10~km~s$^{-1}$ 
to evolve past the MS turn-off point in the HR diagram. 
In general, rotation seems to be relevant even in the case of post-MS, yellow supergiant stars (YSGs). 
Neugent et al. (2012) show that models with rotation agree better with the observed properties of YSGs 
in the LMC than do models with no rotation.

These results illustrate the importance of rotationally-induced mixing leading to chemically quasi-homogeneous
evolution of low mass secondaries in binary systems that are synchronized.
Rotationally-induced mixing may be fundamental in understanding
observational features of some CVs, black hole binaries or Type Ia SNe and to even some single solar-type stars. This possibility
deserves further consideration and modeling.

We thank the MESA team for making this valuable tool readily available
and especially thank Bill Paxton for his ready advice and council 
in running the code. EC would like to thank JJ Hermes for useful discussions.
This research is supported in part by NSF AST-1109801.


{}                     

\newpage

\appendix

\section{ROTATIONAL MIXING IN MESA: COMPARISON WITH OTHER CODES AND WITH OBSERVATIONS}

The effects of rotation in the transport of angular momentum and chemical mixing were only
recently implemented in the stellar evolution code MESA (Paxton et al. 2011) that we are using in the present paper.
We present here comparisons of MESA with other established codes. 
Specifically, we focus on the efficiency of the Spruit-Tayler (ST) mechanism (Spruit 1999, 2002) when the effects of magnetic fields are
taken into account. While its precise utility and implementation may be questioned, the effect of the magnetic torques on angular momentum 
transport as implemented by Spruit is used in many modern stellar evolution codes. 
The effects of the magnetic field on chemical mixing of elements is even more the subject of debate
(Maeder \& Meynet 2003; 2004; 2005, Spruit 2006).  In some calculations chemical mixing by the ST mechanism is 
found to lead to overly efficient mixing and surface N abundances
that are higher than those suggested by observations of massive early B type stars in the Milky Way and in the LMC and SMC
(Hunter et al. 2008; 2009). We therefore also focus on assessing the degree to which MESA models with rotation
and magnetic fields agree with observations. We stress that this comparison is restricted to massive ($>$~10~$M_{\odot}$)
stars since we are not aware of observations of rapidly rotating solar mass stars and relative N abundance measurements. We note, however,
that the findings of the present paper are observationally motivated (depleted carbon abundaces in rapidly rotating secondary stars in
close binaries) and might constitute the first attempt to compare the effects of rotation and magnetic fields in solar type stars 
with direct observations. 

The effects of rotation on angular momentum transport (but not chemical mixing), including magnetic torques as parametrized by Spruit,
have been included in several stellar evolution codes in the past. For the purposes of benchmarking
MESA against some of those well-established codes we ran models similar to those presented by Brott et al. (2011a; b), Ekstr{\"o}m et al. (2012),
Heger, Woosley \& Spruit (2005), Yoon et al. (2006) and Yoon, Dierks \& Langer (2012). The models we have selected have been computed by three
different stellar evolution codes that include the effects of rotation and, in some cases, magnetic fields: the Geneva code 
(Eggenberger et al. 2008), the Yoon \& Langer (2005) code and an updated version of the KEPLER code (Weaver et al. 1978) that was first
used by Heger, Woosley \& Spruit (2005) to study progenitors of Gamma-Ray Bursts (GRBs). 
The majority of the models computed in these works are for massive stars. Therefore most of our comparisons
are for massive stars with the exception of solar mass models computed in the grid of Ekstr{\"o}m et al. (2012). For all models run
with MESA the same parameters (initial mass and metallicity, ZAMS rotational velocity, mixing and angular momentum transport efficiency) 
have been adopted as the models with which we compare. The efficiency of rotational mixing ($f_{c}$) is the main uncertain parameter for
all models calculated. Its value is mainly calibrated by observations of nearby rotating stars. The choice of 0.046 was proposed by
Pinsonneault et al. (1989) in order to explain solar lithium abundances, while theoretical predictions by Chaboyer \& Zahn (1992) provided
a somewhat different value of 0.033 which was adopted in calculations of massive rotating models done by Heger et al. (2000). A somewhat
smaller value of 0.0228 calibrated against observations of massive B type stars in the LMC was adopted by Brott et al. (2011a; b). 

We first computed 10~$M_{\odot}$ models for $Z =$~0.0047 (the metallicity of the LMC) for four different equatorial rotational velocities
at the ZAMS: 58, 116, 232 and 345~km~s$^{-1}$ and for two cases: a case without the effects of ST on the chemical mixing included and a second
case that includes those effects. Those parameters are exactly the same as those in Brott et al. (2011b). Specifically
we compare their result on the evolution of surface nitrogen enrichment presented in their Figure 8 with our results, presented in Figure A1. 
As can be seen, the inclusion of the ST mechanism does lead to somewhat higher surface nitrogen abundances than when it is not included,
but in both cases the values we find for 12+$\log[N/H]$ are consistent with those found by Brott et al. (2011b). 
Interestingly, the results between the two codes agree much better when ST chemical mixing is included. 

Next, we consider a 1~$M_{\odot}$ model at solar metallicity ($Z =$~0.014) for a non-rotating case and a case with a rotational velocity
of 50~km~s$^{-1}$. The effects of ST are accounted for in these models only in the transport of angular momentum and not in chemical mixing. 
In Figure A2 we plot the evolution of the surface helium abundance (upper left panel) as well as the evolution
of the surface abundance ratios [N/C] (upper right panel) and [N/O] (lower left panel) for the non-rotating (black curve) and the rotating
(red curve) case. The end of the H-burning phase is indicated with the dashed vertical line in all panels. 
This figure can be directly compared to the results presented in Table 2 of Ekstr{\"o}m et al. (2012) for the same models. We see that the
values for the surface He abundance we estimate with MESA at the end of the H-burning phase are generally a little higher (0.02-0.05 in mass fraction) 
than the ones presented by Ekstr{\"o}m et al., but the surface abundance ratios are very similar in the two works. Any differences may be
attributable to different algorithms between the codes and to numerical errors (due to differences in resolution, for example).

For another comparison of MESA including the effects of rotation and of the ST mechanism accounting only for the angular momentum transport
we choose models recently presented by Yoon, Dierks \& Langer (2012) in their investigation of the evolution
of massive population III stars. Specifically, we computed a 10~$M_{\odot}$ model of zero metallicity rotating at 40\% the critical rotational
velocity at the ZAMS. For this model we show the MESA-computed H-R diagram in Figure A3 that can be directly compared to the one presented
by Yoon, Dierks \& Langer (2012) in their Figure 1 (upper left panel). MESA gives similar values and ranges for the effective temperature,
$T_{eff}$, and the luminosity except that at later stages MESA gives slightly higher luminosities. 
We note that some peculiar ``kinks" on the HR track for this model are numerical artifacts related to 
limited resolution. Similar numerical features are also seen in calculations of rotating progenitors of GRBs by Yoon \& Langer (2005) (see
the HR tracks presented in their Figure 2).
Also,  at the
end of the calculation MESA gives carbon-oxygen (C/O) core mass of 1.72~$M_{\odot}$ while Yoon, Dierks \& Langer (2012) find it to be 1.78~$M_{\odot}$ 
and a surface helium mass fraction of 0.348 in contrast with 0.32 that they find. 

Now we proceed to benchmark MESA against other available results that include the effects of the magnetic fields both in angular momentum
transport and in mixing of chemical elements. For these comparisons we start by considering the rotating, magnetic 15~$M_{\odot}$ model at
solar metallicity that was presented by Heger, Woosley \& Spruit (2005). This is the archetypical paper that was
used to implement rotation and magnetic fields in MESA. One of the MESA built-in test problems is the same as the one
we present here, and it is the original model that was used to evaluate the effects of rotation as calculated by MESA. 
For this model, an initial ZAMS rotational velocity of 200~km~s$^{-1}$ is used. Figure A4 shows the evolution of He, C, N and O surface
mass fractions as calculated by MESA for a non-rotating (black curves), a rotating (red curves) and a rotating plus magnetic (blue curves) case. 
The dashed vertical lines of the same colors represent the time at which the central hydrogen mass fraction was down to $\sim$~35\% for each model
in order to directly compare with the results presented in Table 6 of Heger, Woosley \& Spruit (2005). We see very good agreement of the MESA results
with the results of Heger, Woosley \& Spruit (2005). We note that the inclusion of chemical mixing due to ST leads to somewhat larger abundance changes and, 
consequently, stronger surface nitrogen enhancements. This result is also obtained by Maeder \& Meynet (2004) were they suggest
that such differences might indicate that the Spruit (2002) prescriptions for chemical mixing due to magnetic fields lead to overestimates. 

The other available independent calculations of stellar evolution with rotation and magnetic fields in both angular momentum transport and
chemical mixing are found in Yoon et al. (2006). To perform a comparison with MESA we select the 12~$M_{\odot}$ model with 
initial metallicity of $Z =$~0.004 and initial (ZAMS) equatorial rotational velocity of 354.14~km~s$^{-1}$. In Figure A5 we show the
evolution of the surface nitrogen abundance for this model as calculated by MESA in the case for which ST is not considered in chemical
mixing (black curve) and in the case for which it is (red curve). At time $t =$~$1.93 \times 10^{7}$~years, we see that 12+$\log[N/H]$~$\simeq$~8.1-8.4
for the two cases. Using the results presented in Table 4 of Yoon et al. (2006) we estimate that in their model 12+$\log[N/H] =$~8.26, a value
that is in good agreement with ours.
At the same time in the evolution, the mass of the C/O core is 1.91~$M_{\odot}$
for the MESA model that includes the effects of ST on mixing. For the same model calculated by Yoon et al. (2006) the mass of the
C/O core is 1.892~$M_{\odot}$, indicating that the mixing processes that lead to the formation of the core in the two codes also give
very similar result at the same evolutionary stage. 

The comparison of MESA stellar evolution with rotation and magnetic fields to these other codes indicates
that MESA accounts for these effects as well as they do, specifically if one accounts for differences between codes, 
numerical errors, different model resolutions
and specific numerical techniques that can interact in non-linear ways. Major indicators of
the effects of rotation and magnetic fields on chemical mixing such as surface nitrogen and helium enrichements and C/O core masses 
are found to have a similar range of values amongst different stellar evolution codes.

This agreement enables us to take a step further and directly compare the MESA results on surface nitrogen abundances of rotating stars
with spectroscopic observations of rotating stars in the Milky Way, the LMC and the SMC recently presented by the
VLT-FLAMES survey of massive stars (Hunter et al. 2008, 2009). To enable this comparison, we computed an additional MESA model for a 15~$M_{\odot}$
star at solar metallicity ($Z =$~0.014) rotating at 150~km~s$^{-1}$
for two cases: one with the effects of the ST mechanism on chemical mixing ignored and one with those effects
considered. The evolution of 12+$\log[N/H]$ for this model is presented in Figure A6. Direct comparison with Figure 6 of
Hunter et al. (2009) illustrates that MESA estimates for the range of 12+$\log[N/H]$ values are consistent with those measured
for massive rotating stars in the Milky Way. Similar results are obtained for the same model at the higher rotational velocity
of 200~km~s$^{-1}$, which we calculated in order to compare with the results of Heger, Woosley \& Spruit (2005). The higher
rotational velocity led to slightly higher values for 12+$\log[N/H]$ but still within the range of the observations presented
by Hunter et al. (2009). 
In addition, the models that we calculated for a metallicity of $Z =$~0.0047 in order to compare with the results
of Brott et al. (2011b) show 12+$\log[N/H]$ values (Figure A1) that agree very well with the observations of Hunter et al. (2009) for massive
early B type stars in the LMC (their Figure 5). 

As we mentioned above, there are currently no existing 12+$\log[N/H]$ 
observations in single solar type rapidly rotating stars 
that can be compared to evolutionary model predictions in the low mass range. Just such a comparison
was the initial
observational motivation of the present work: to account for carbon depletion in the secondary stars of some CVs and LMXBs which are 
close binaries and tidally locked, therefore rapidly rotating, based on calibrations
that have been done for higher mass stars (and for the sun) for which observations are available.
There certainly is a degree of uncertainty associated with the calculations, specifically
when the effects of the ST mechanism on chemical mixing are included. In general the inclusion of these effects leads to somewhat higher
abundances for some models of massive rapidly rotating stars  
than those suggested by some observations while in some calculations the results are consistent with observations (Maeder \& Meynet 2005). 
Given the differences in computational algorithm that were discussed above 
and the uncertainties in the measurement of initial metallicities (upon which the results for surface abundance changes are sensitive) it remains
unclear whether the inclusion of the effects of ST mixing necessarily lead to wrong results. In the cases investigated by this paper it is
suggested that those effects of increased mixing efficiency due to the ST mechanism are necessary if rotation and magnetic fields can be
used as an explanation. To illustrate the importance of ST mixing in our models we present the distribution of the diffusion coefficients
due to meridional circulation (Eddington-Sweet circulation; ES), the Goldreich-Schubert-Fricke instability (GSF) and the ST mechanism 
in Figure A7 for one of our models (1~$M_{\odot}$ rotating at 30\% the critical value) at about the half of its main sequence 
lifetime for a case where ST was ignored in calculations of the chemical mixing (left panel) and a case where ST was included (right panel).
From this plot it is obvious that ST dominates chemical mixing through most of the stellar interior as well as the surface,
and in some cases it is more than an order of magnitude more important than meridional circulation. 


\begin{figure}       
\centerline{
\hskip -0.05 in
\psfig{figure=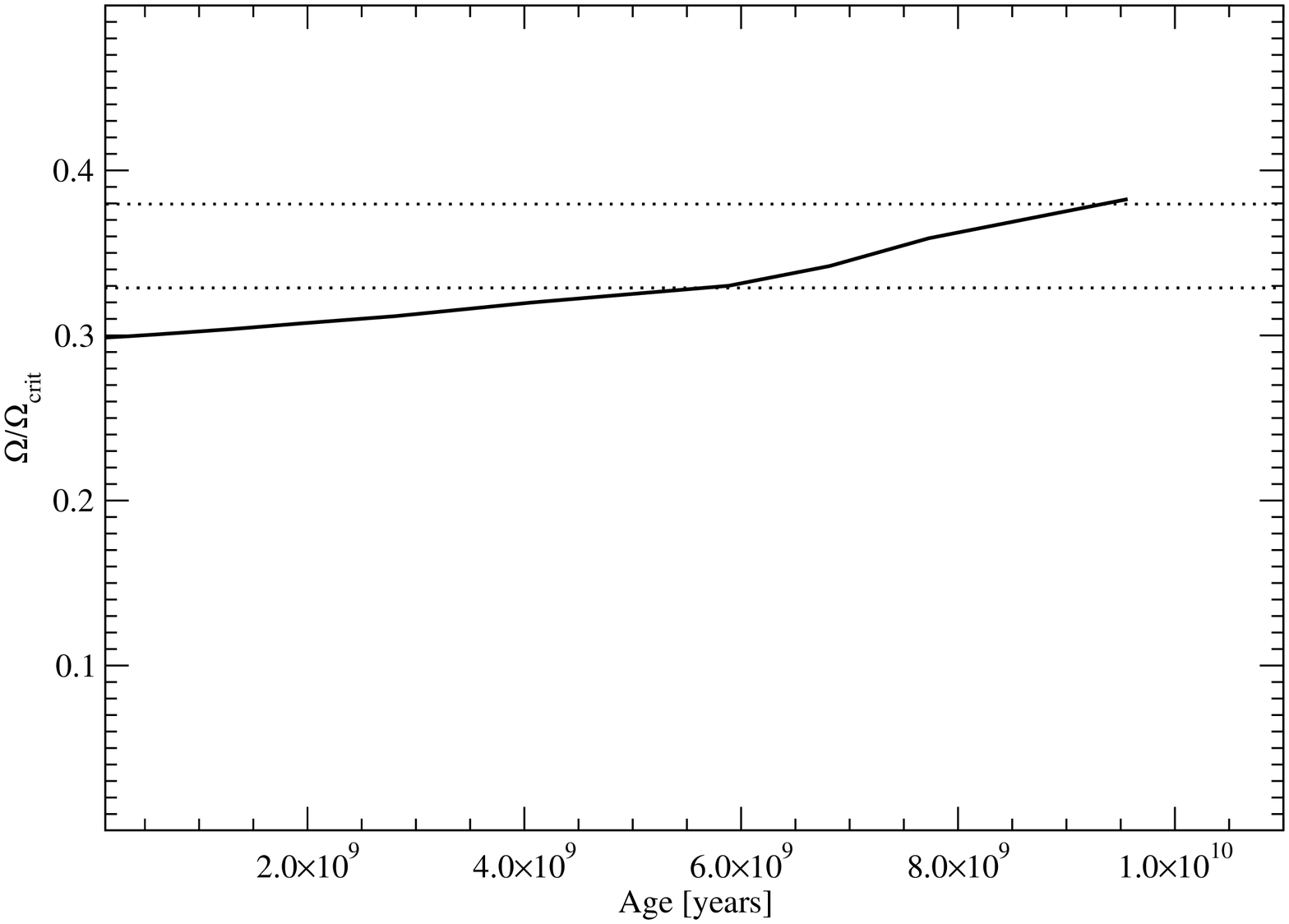,angle=-90,width=3.5in}
\hskip -0.05 in
\psfig{figure=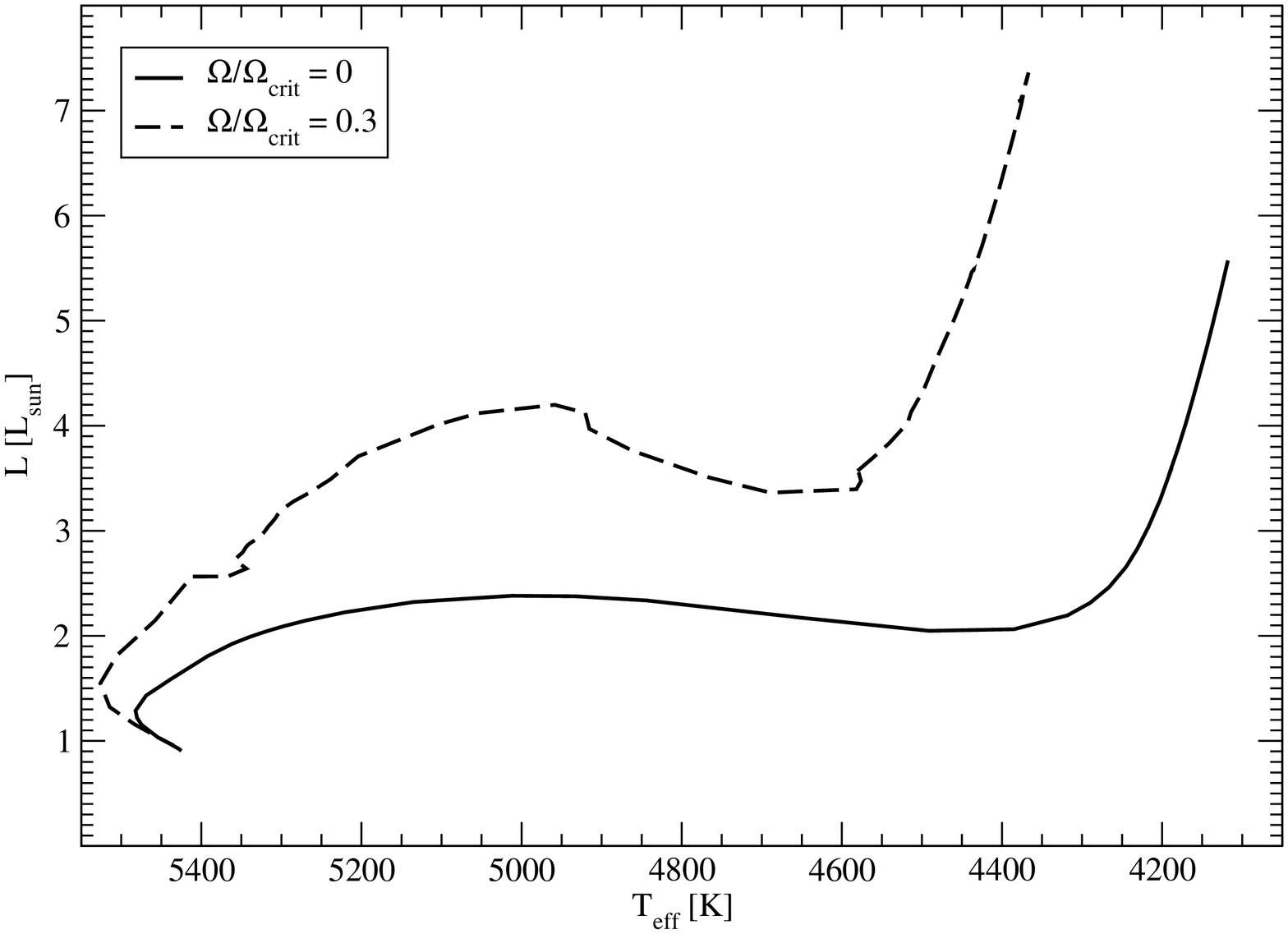,angle=-90,width=3.5in}
}
\caption{{\it Left Panel}: Evolution of $\Omega/\Omega_{crit}$, during the MS for the 1~$M_{\odot}$, $Z =$~$Z_{\odot}$,
$\Omega/\Omega_{crit} =$~0.3 (solid curve) model. The two horizontal vertical lines indicate the range of $\Omega/\Omega_{crit}$
for which the secondary star fills its Roche lobe (see \S2). 
{\it Right Panel}: Evolution of the 1~$M_{\odot}$, $Z =$~$Z_{\odot}$, $\Omega/\Omega_{crit} =$~0 (solid curve) and
$\Omega/\Omega_{crit} =$~0.3 (dashed curve) models in the H-R diagram.}
\end{figure}

\begin{figure}
\begin{center}
\includegraphics[angle=-90,width=20cm]{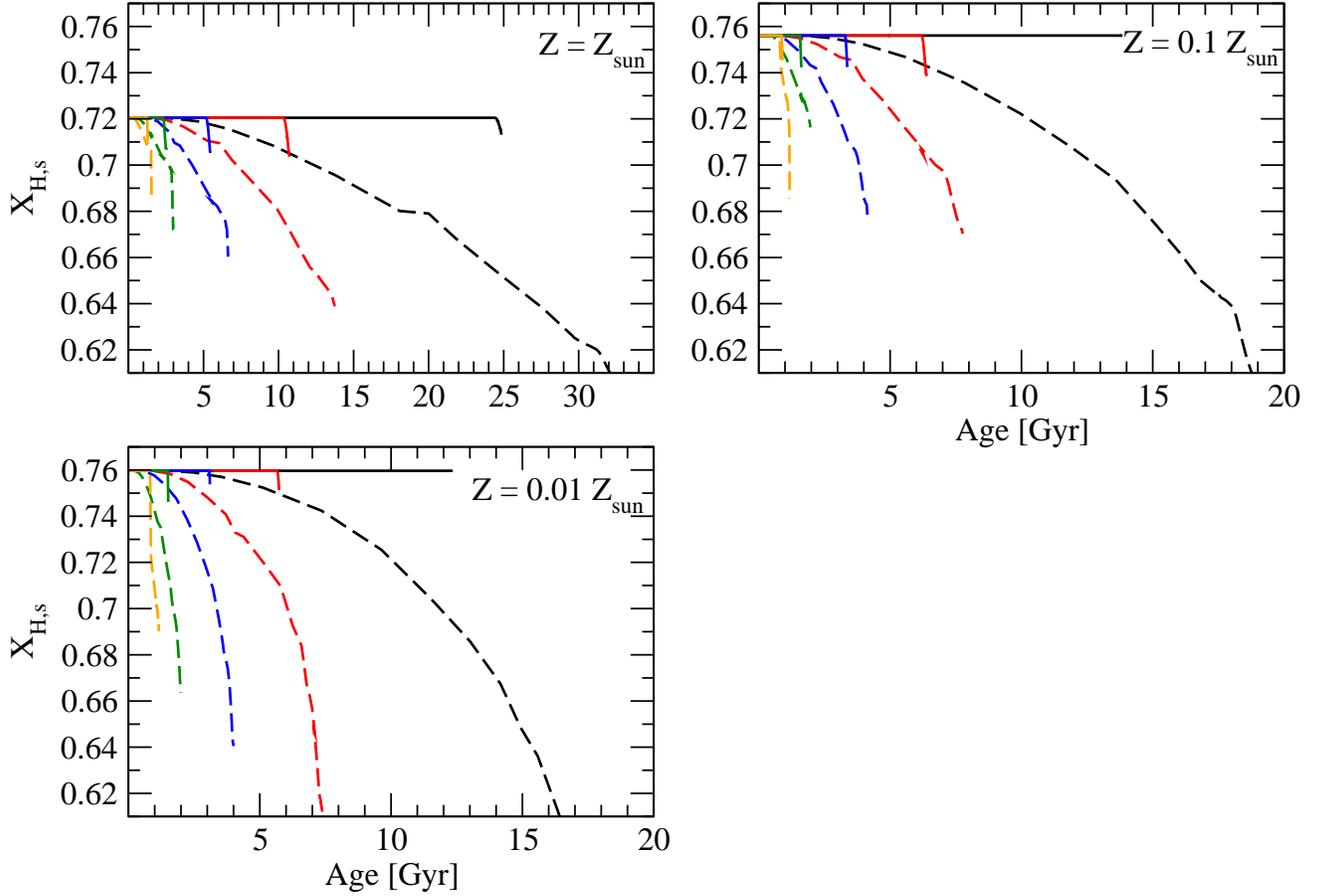}
\caption{Evolution of the surface hydrogen mass fraction, $X_{H,s}$, for the 0.8~$M_{\odot}$ (black curves),
1~$M_{\odot}$ (red curves), 1.2~$M_{\odot}$ (blue curves), 1.5~$M_{\odot}$ (green curves) and 1.8~$M_{\odot}$ (orange curves)
models with $Z =$~$Z_{\odot}$ (upper left panel), $Z =$~0.1~$Z_{\odot}$ (upper right panel) and $Z =$~0.01~$Z_{\odot}$ (lower left panel).
The solid curves are for $\Omega/\Omega_{crit} =$~0 and the dashed
curves for $\Omega/\Omega_{crit} =$~0.3. The steep vertical lines for the non-rotating models represent the onset of 
evolution to the red giant branch, as shown in Figure 6.}
\end{center}
\end{figure}

\begin{figure}
\begin{center}
\includegraphics[angle=-90,width=20cm]{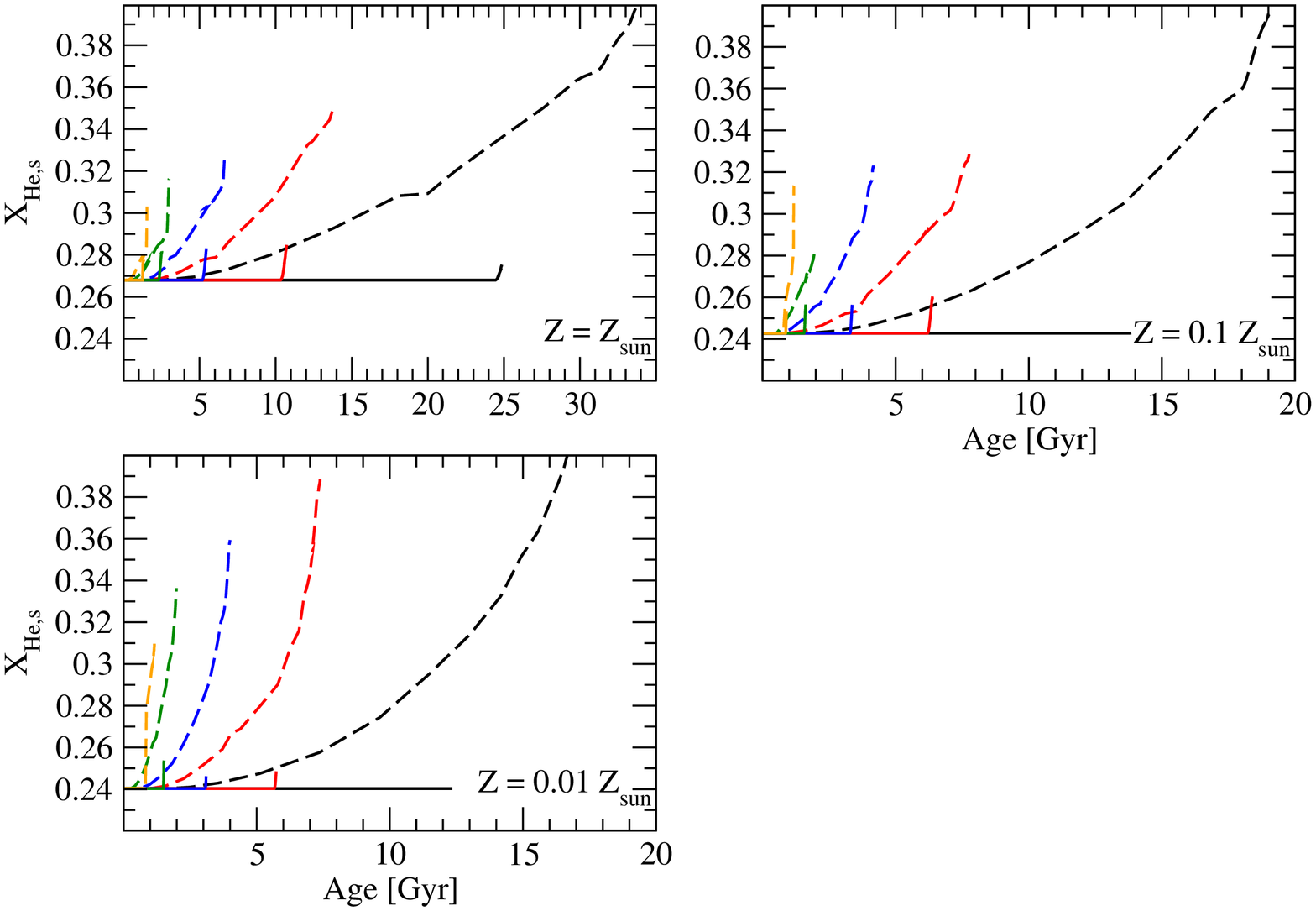}
\caption{Evolution of the surface helium mass fraction, $X_{He,s}$. The curves represent the same models as those
decribed in the caption of Figure 2.}
\end{center}
\end{figure}

\begin{figure}
\begin{center}
\includegraphics[angle=-90,width=20cm]{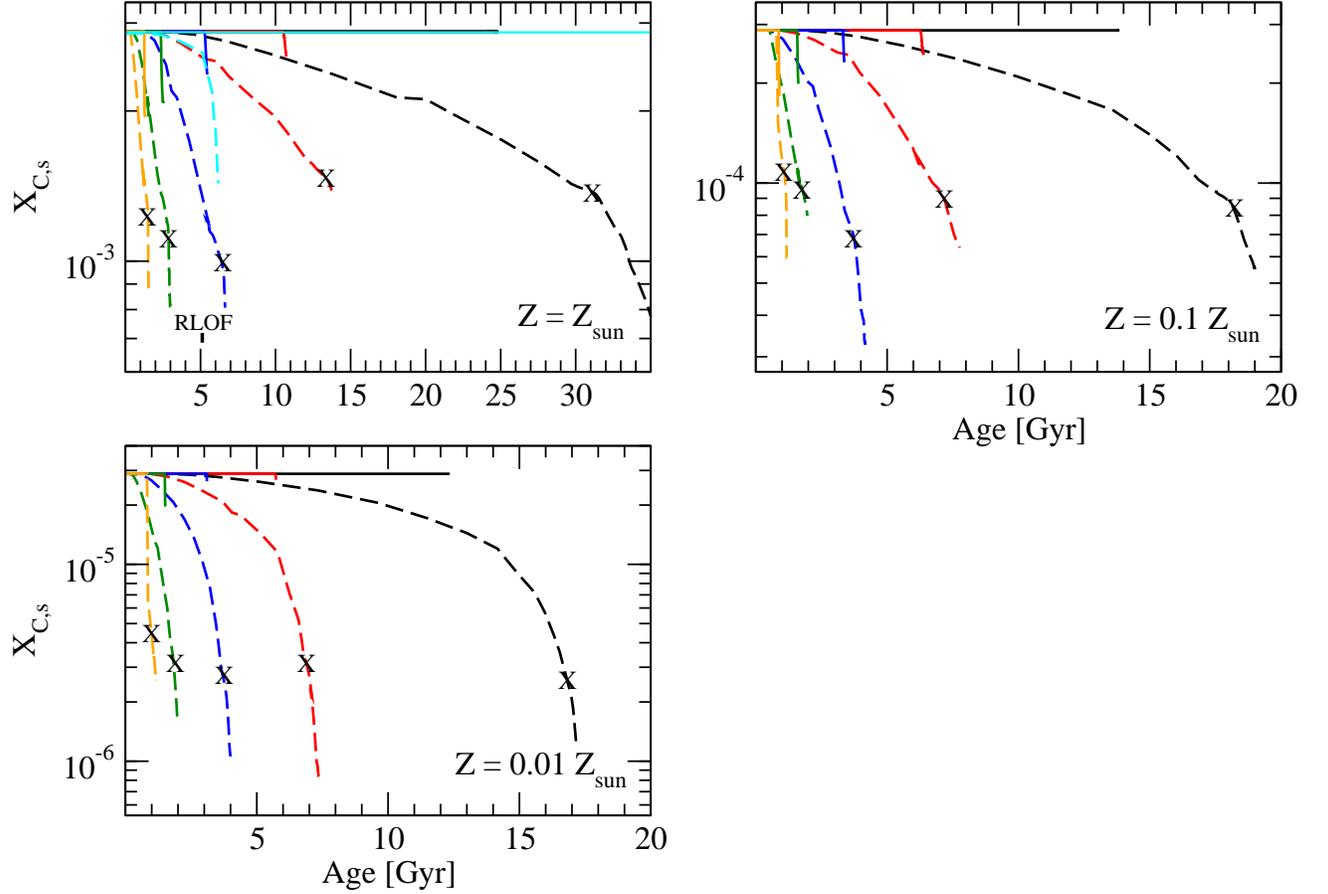}
\caption{Evolution of the surface carbon mass fraction, $X_{C,s}$. The curves represent the same models as those
decribed in the caption of Figure 2.
The light blue curves show the evolution of $X_{C,s}$ for the
1~$M_{\odot}$, $Z =$~$Z_{\odot}$, $\Omega/\Omega_{crit} =$~0 (solid) and the
1~$M_{\odot}$, $Z =$~$Z_{\odot}$, $\Omega/\Omega_{crit} =$~0.3 (dashed)
models in the case where we adopt a RLOF mass loss rate
$\dot{M}_{RLOF} =$~$5 \times 10^{-10}$~$M_{\odot}$~yr$^{-1}$ starting at $t_{RL} =$~$5 \times 10^{9}$~years The
``X" marks indicate the end of the MS for the rotating models. The time of RLOF mass loss is also indicated.}
\end{center}
\end{figure}

\begin{figure}
\begin{center}
\includegraphics[angle=-90,width=20cm]{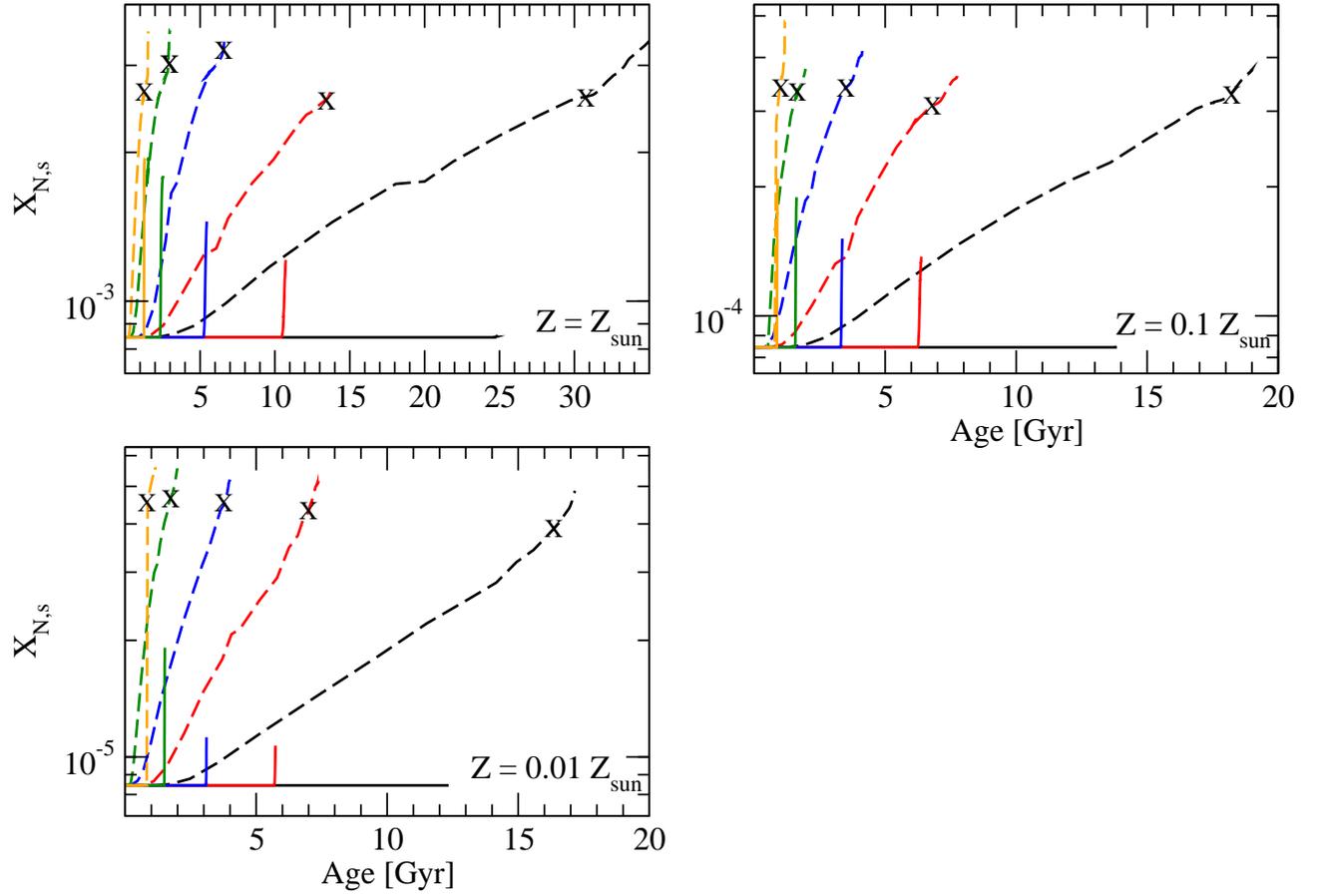}
\caption{Evolution of the surface nitrogen mass fraction, $X_{N,s}$. The curves represent the same models as those
decribed in the caption of Figure 2.}
\end{center}
\end{figure}

\begin{figure}
\begin{center}
\includegraphics[angle=-90,width=20cm]{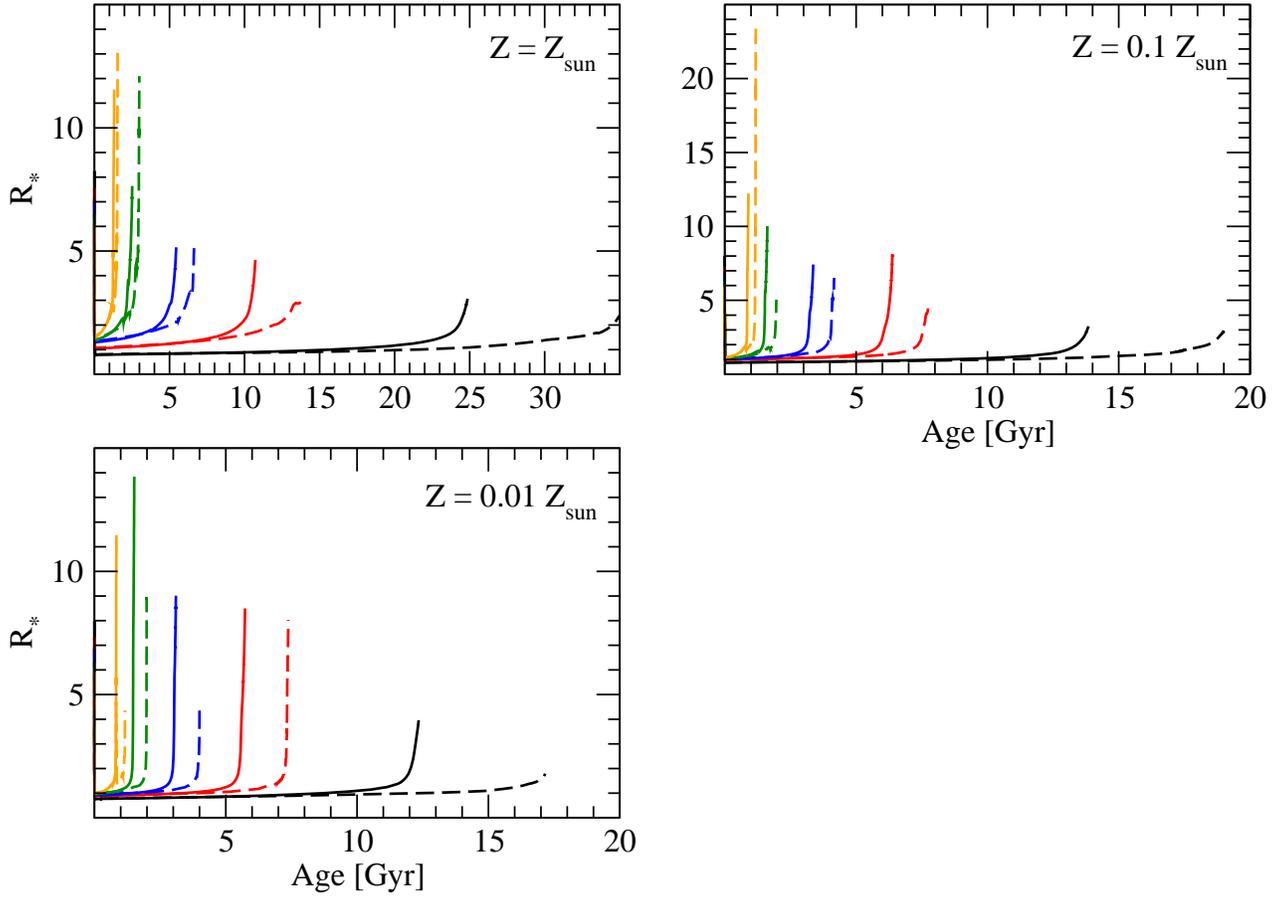}
\caption{Evolution of the radius of the secondary star, $R_{*}$. The curves represent the same models as those
decribed in the caption of Figure 2.}
\end{center}
\end{figure}

\begin{figure}
\begin{center}
\includegraphics[angle=-90,width=20cm]{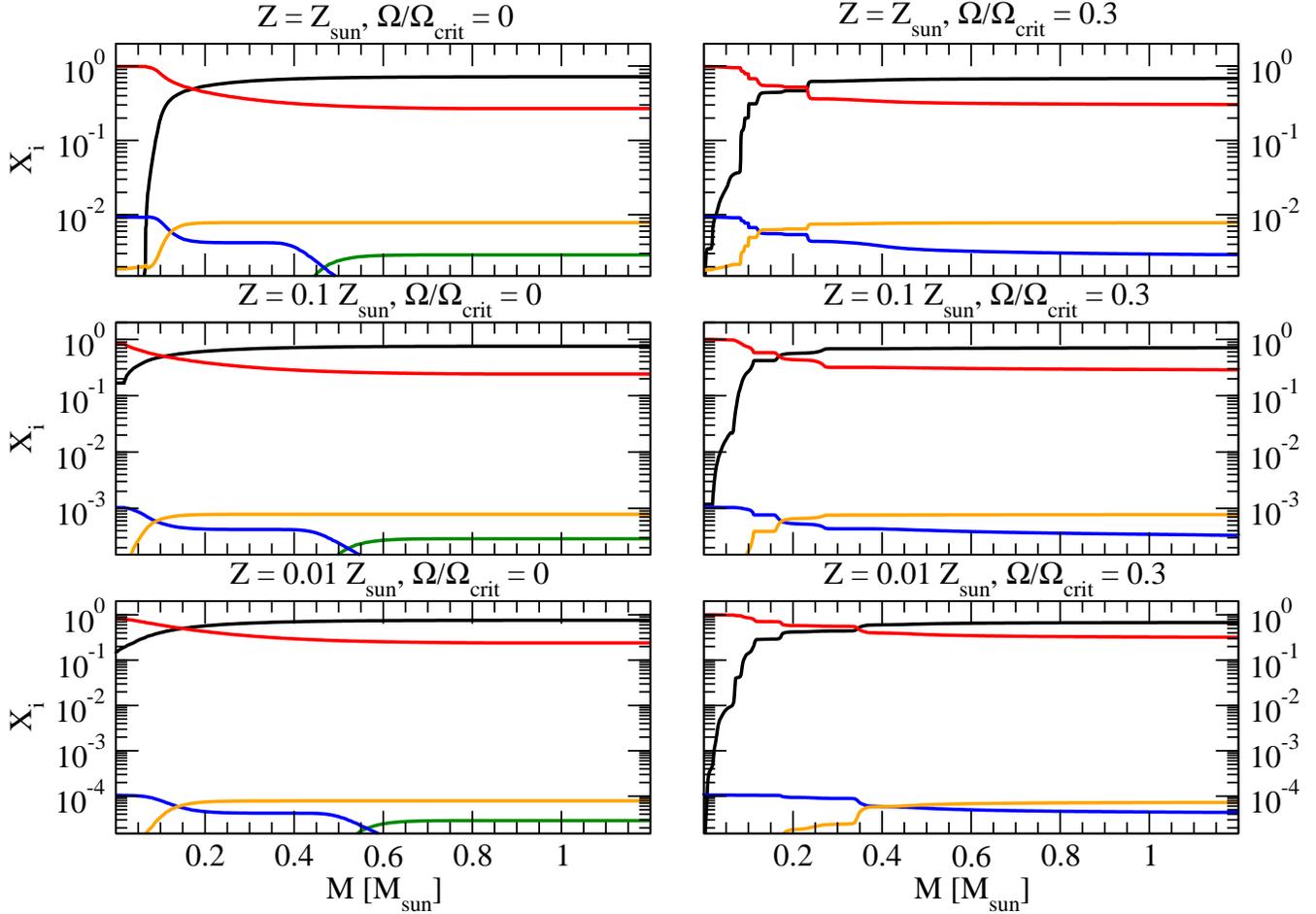}
\caption{Composition structures of the 1.2~$M_{\odot}$, $Z =$~$Z_{\odot}$, $\Omega/\Omega_{crit} =$~0 and $\Omega/\Omega_{crit} =$~0.3 (upper left
and upper right panel respectively) models, the 1.2~$M_{\odot}$, $Z =$~0.1~$Z_{\odot}$, $\Omega/\Omega_{crit} =$~0 and $\Omega/\Omega_{crit} =$~0.3 (middle left
and middle right panel respectively) models and the 1.2~$M_{\odot}$, $Z =$~0.01~$Z_{\odot}$, $\Omega/\Omega_{crit} =$~0 
and $\Omega/\Omega_{crit} =$~0.3 (lower left
and lower right panel respectively) models at time $\simeq$~3/4~$\tau_{MS}$. 
In all panels the mass fractions of the following elements are plotted: H (black curves), He (red curves),
C (green curves), N (blue curves) and O (orange curves). The discrete steps in the H and He profiles for the rotating models are due to mixing resulting
from the rapidly changing magnetic field, and therefore, magnetic viscosity, in these regions.}
\end{center}
\end{figure}

\begin{figure}       
\centerline{
\hskip -0.05 in
\psfig{figure=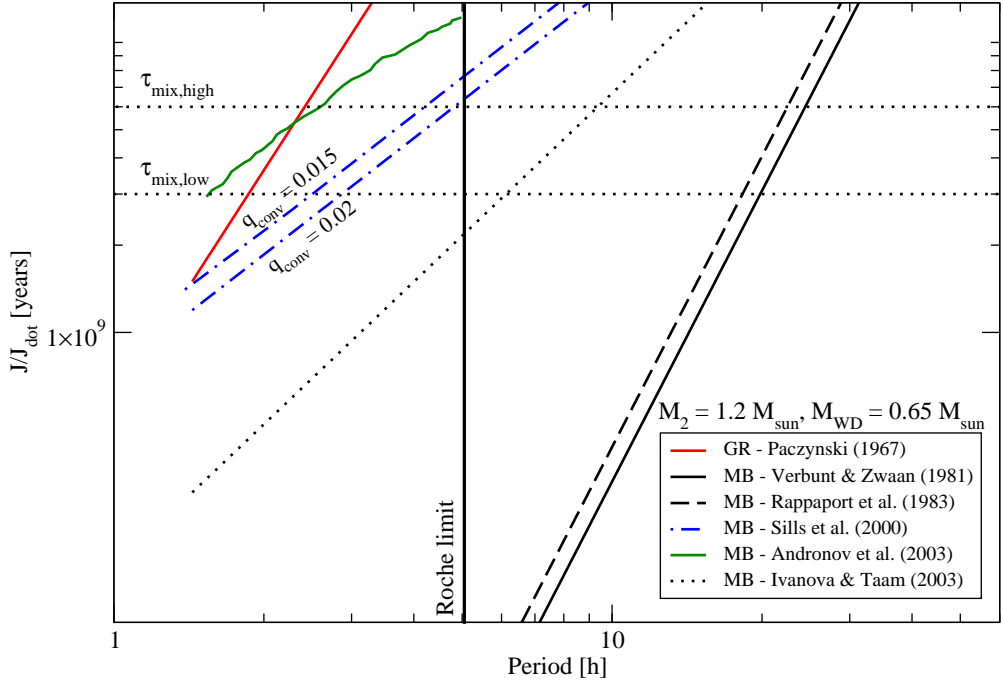,angle=-90,width=6in}
}
\caption{Characteristic time-scales for angular momentum loss for a range of orbital periods in the case
of a system with a secondary star of mass $M_{2} =$~1.2~$M_{\odot}$ and a white dwarf of mass $M_{WD} =$~0.65~$M_{\odot}$, characteristic
of SS Cyg. The
solid red curve represents the angular momentum loss time-scale due to gravitational radiation from Paczynski (1967). The solid
and dashed black curves represent $\tau_{MB}$ as calculated by Verbunt \& Zwaan (1981) and Rappaport et al. (1983) for $\gamma=0$
respectively. The dot-dashed blue curves show the longer magnetic braking time-scale according to the predictions of Sills et al. (2000)
for $q_{conv} =$~0.015 and $q_{conv} =$~0.02. The dotted black curve is based on Ivanova \& Taam (2003). The solid green curve
is data for the numerically calculated angular momentum loss time-scale taken from the upper panel of Figure 4 of Andronov et al. (2003)
in the case of an evolved secondary. The horizontal dotted lines indicate the characteristic time-scales required for significant rotationally-induced 
mixing to occur based on the results of our evolutionary calculations in \S 3.1. Finally, the solid vertical line indicates the Roche limit
expressed in terms of orbital period for this particular system.}
\end{figure}

\begin{figure}       
\centerline{
\hskip -0.05 in
\psfig{figure=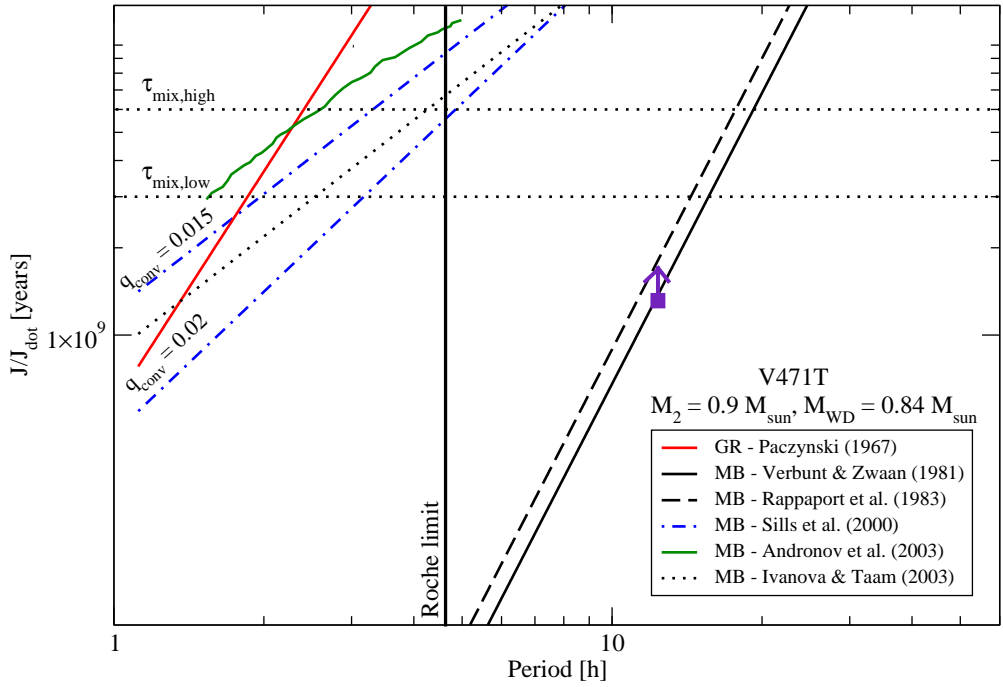,angle=-90,width=6in}
}
\caption{Characteristic time-scales for angular momentum loss for a range of orbital periods in the case
of a system with a secondary star of mass $M_{2} =$~1.2~$M_{\odot}$ and a white dwarf of mass $M_{WD} =$~0.65~$M_{\odot}$, characteristic
of the binary V147~Tau (O'Brien et al. 2001; filled purple square). The curves represent the same models as those
decribed in the caption of Figure 8.}
\end{figure}

\begin{figure}
\begin{center}
\includegraphics[angle=0,width=18cm]{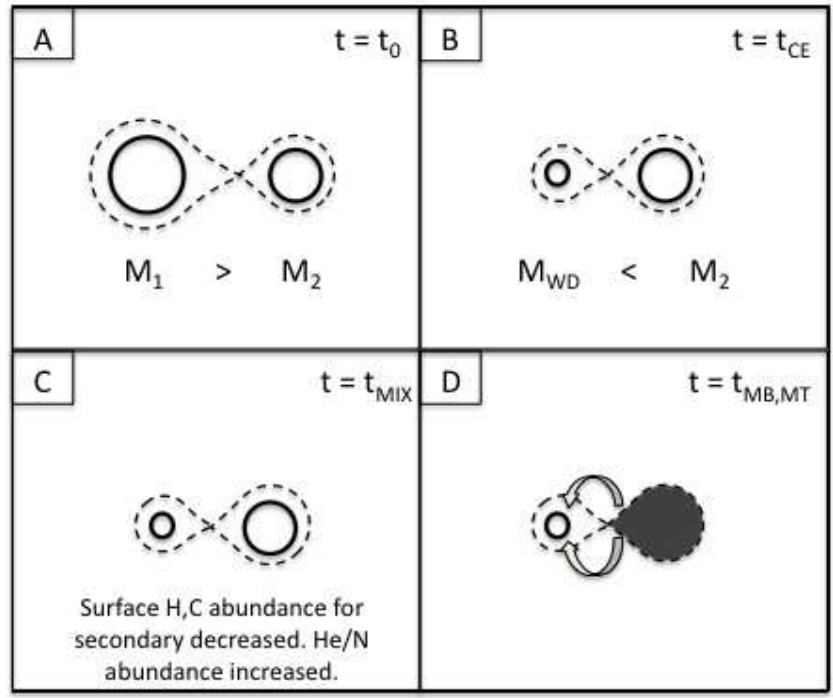}
\caption{Schematic representation of the evolution of a binary system that leads to reduced surface carbon and hydrogen
and enhanced surface helium and nitrogen abundances for the secondary. See text (\S 3.2) for details.}
\end{center}
\end{figure}

\begin{figure}
\begin{center}
\includegraphics[angle=-90,width=20cm]{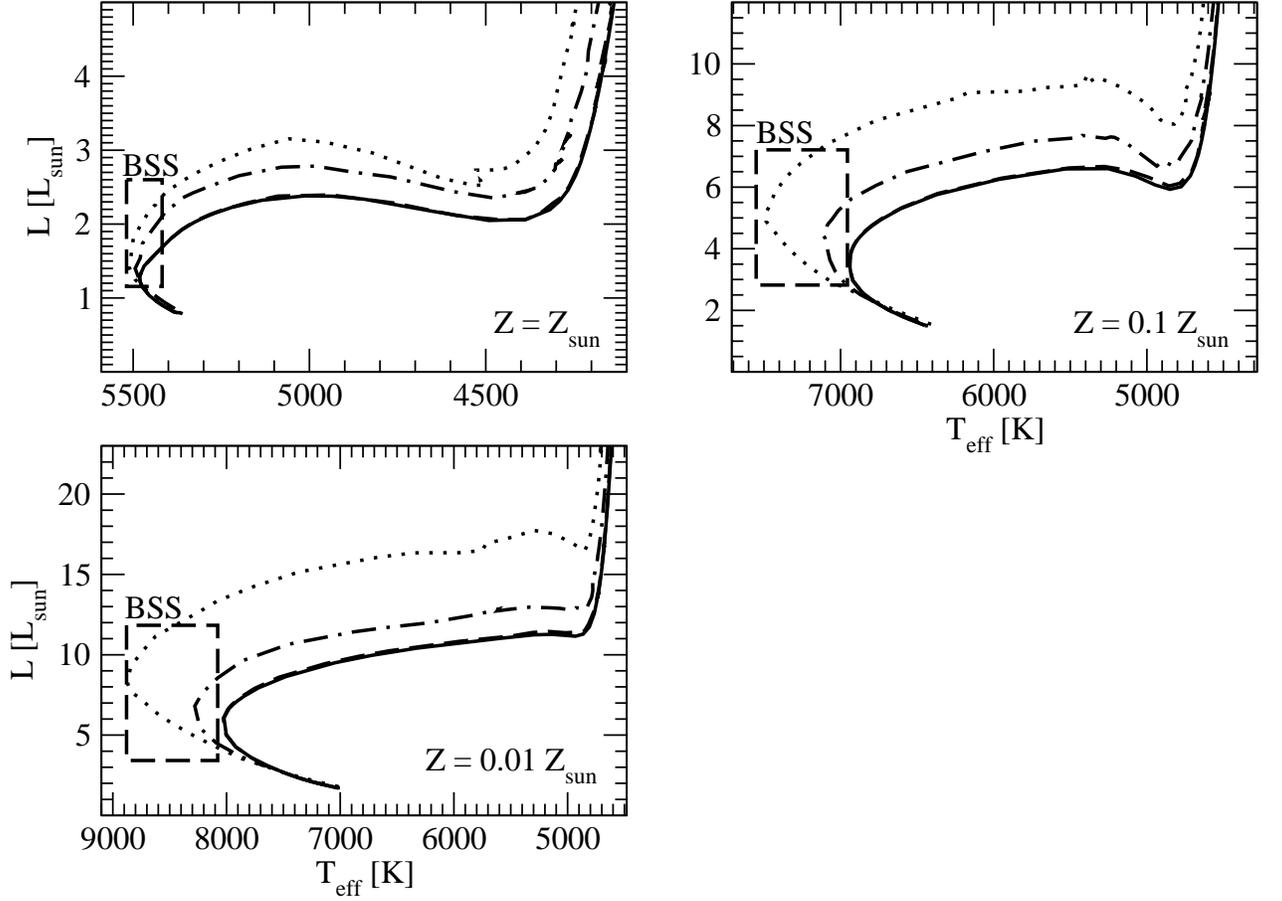}
\caption{Evolution of blue straggler candidate models on the HR diagram for 1~$M_{\odot}$, 
$Z =$~$Z_{\odot}$ (upper left panel), $Z =$~0.1~$Z_{\odot}$ (upper right panel) and $Z =$~0.01~$Z_{\odot}$ (lower left panel).
Solid black curves indicate $v_{rot} =$~0~km~s$^{-1}$, dashed black curves $v_{rot} =$~10~km~s$^{-1}$, dashed-dotted black
curves $v_{rot} =$~60~km~s$^{-1}$ and dotted curves $v_{rot} =$~100~km~s$^{-1}$. 
The boxes marked ``BSS" indicate where the blue straggler stars are expected to be in the HR diagram.}
\end{center}
\end{figure}


\begin{figure}
\begin{center}
\includegraphics[angle=-90,width=18cm]{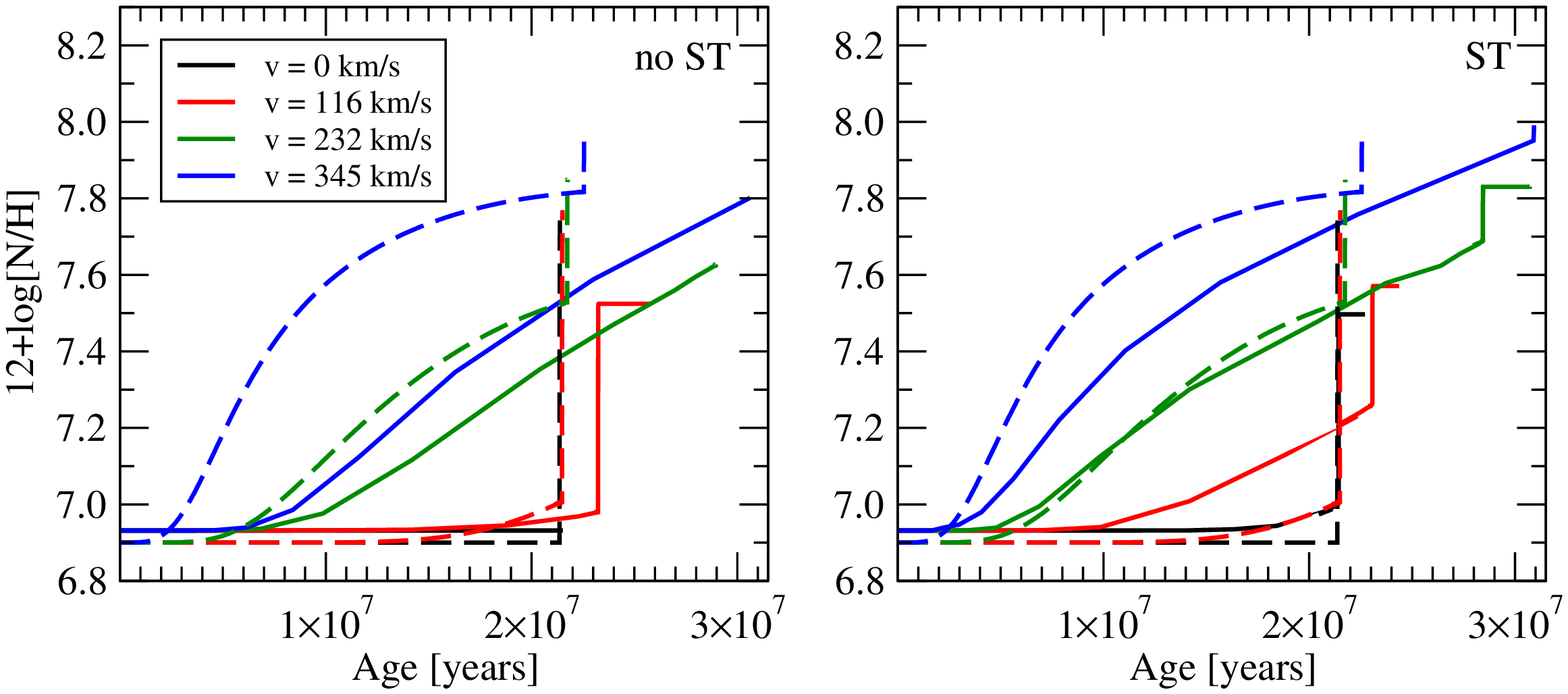}
\renewcommand\thefigure{A1} 
\caption{Evolution of surface nitrogen abundance with time for 10~$M_{\odot}$, $Z =$~0.0047 models for ZAMS rotational velocities of
58 (black curve), 116 (red curve), 232 (green curve) and 345~km~s$^{-1}$ (blue curve) for a case
where the effects of the ST mechanism on chemical mixing are ignored (left panel) and for a case where those effects are included (right panel)
as calculated by MESA. Angular momentum transport via the ST mechanism is included in both cases. The dashed curves represent 
the results from the same models of Brott et al. (2011b).}
\end{center}
\end{figure}

\begin{figure}
\begin{center}
\includegraphics[angle=-90,width=18cm]{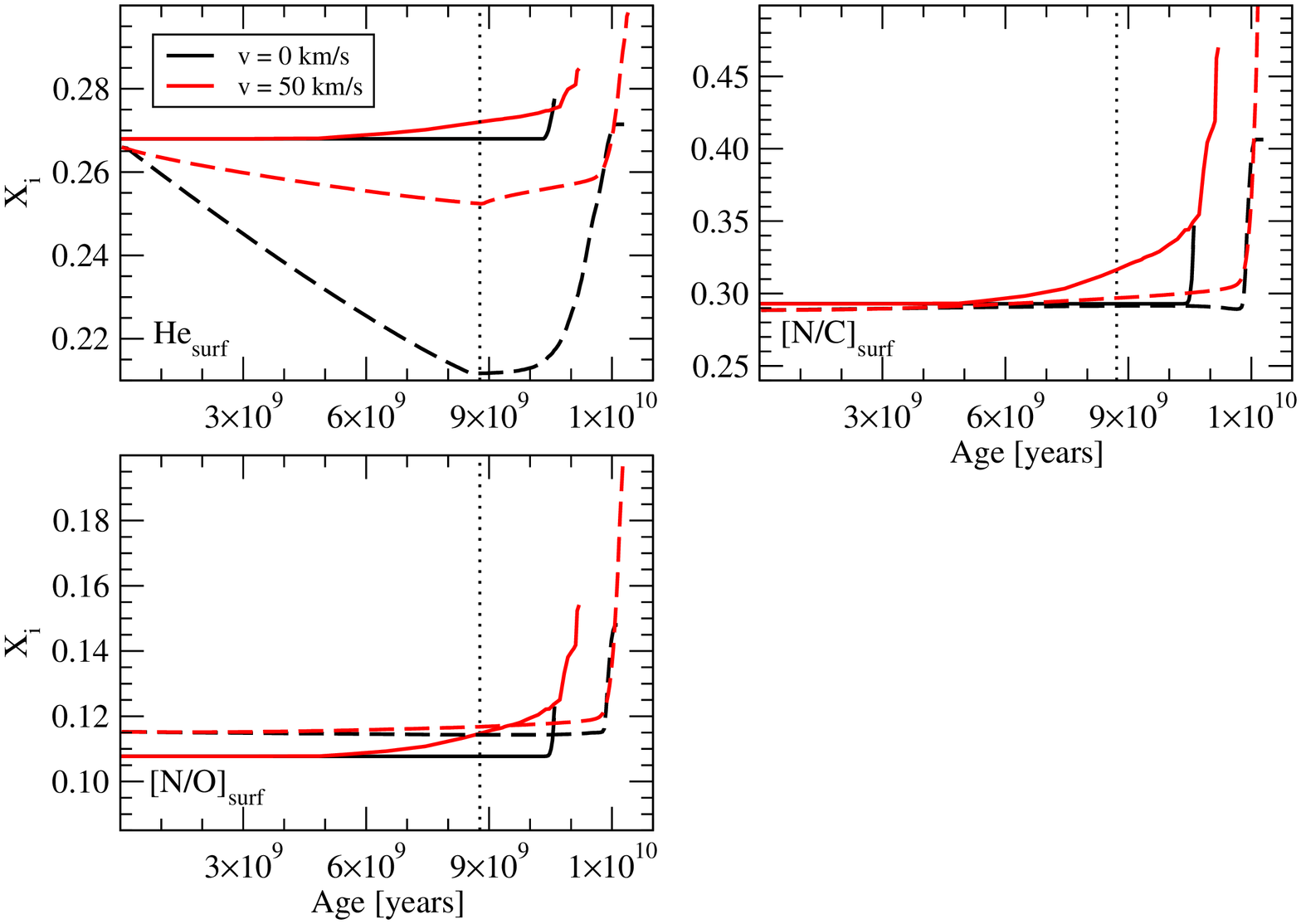}
\renewcommand\thefigure{A2} 
\caption{Evolution of the surface He abundance (upper left panel), the surface [N/C] abundance ratio (upper right panel) and the surface
[N/O] abundance ratio (lower left panel) for a 1~$M_{\odot}$, $Z= $~0.014 non-rotating (black curves) model and a model with initial rotational
velocity of 50~km~s$^{-1}$ (red curves) as calculated by MESA. The dotted vertical lines indicated the end of the H-burning phase for the 
non-rotating models. The effects of the ST mechanism on angular momentum transport were included for this model while its effects on
chemical mixing are not. The dashed curves represent the results from the same models of Ekstr{\"o}m et al. (2012).}
\end{center}
\end{figure}

\begin{figure}
\begin{center}
\includegraphics[angle=-90,width=18cm]{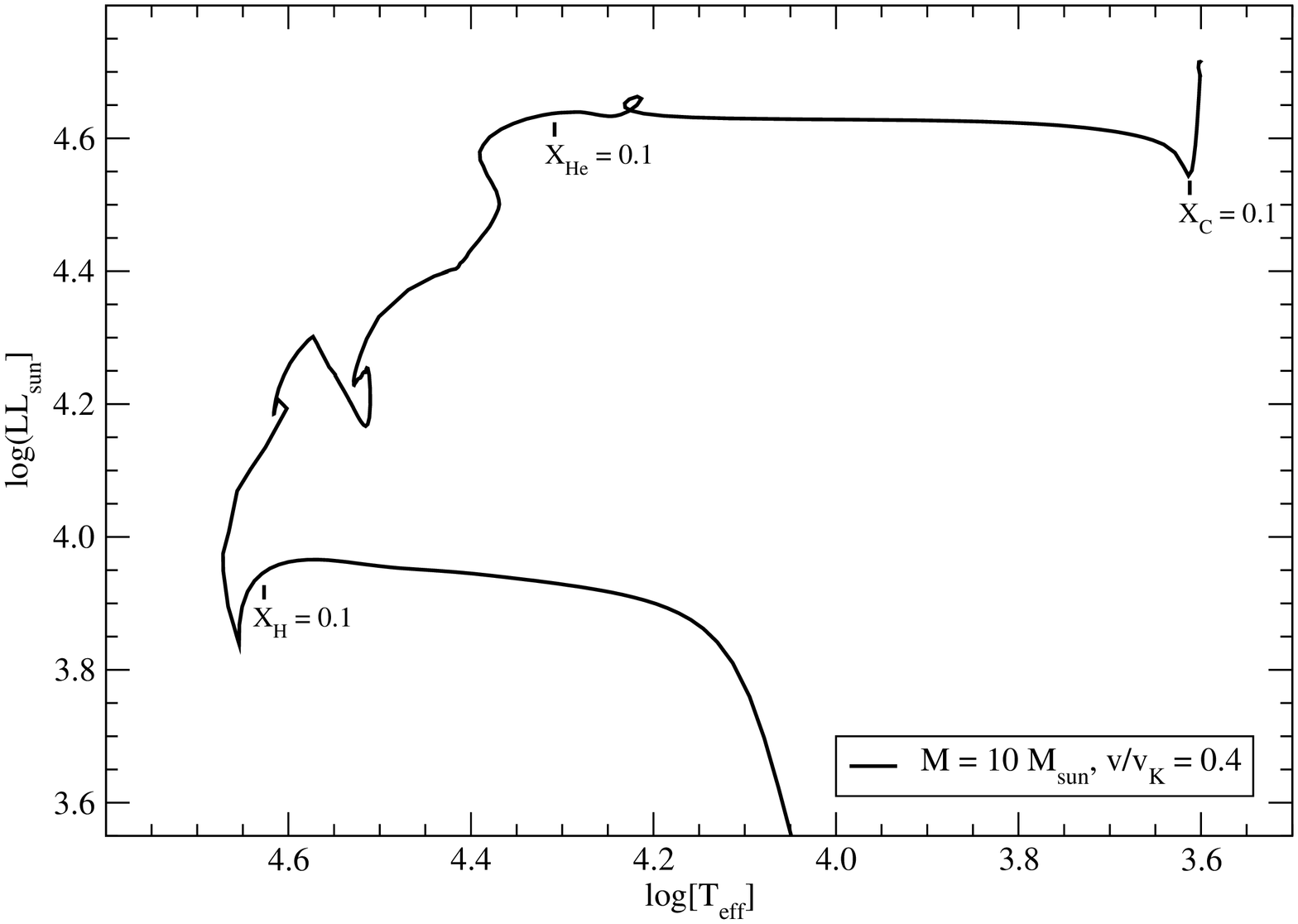}
\renewcommand\thefigure{A3} 
\caption{H-R diagram for a 10~$M_{\odot}$ zero metallicity model rotating at 40\% the critical velocity at ZAMS as calculated in MESA.
The effects of the ST mechanism on angular momentum transport were included for this model, but not its effects on
chemical mixing. The evolutionary stages for which the central mass fractions of hydrogen, helium and carbon are $\simeq$~0.1 are indicated.}
\end{center}
\end{figure}

\begin{figure}
\begin{center}
\includegraphics[angle=-90,width=18cm]{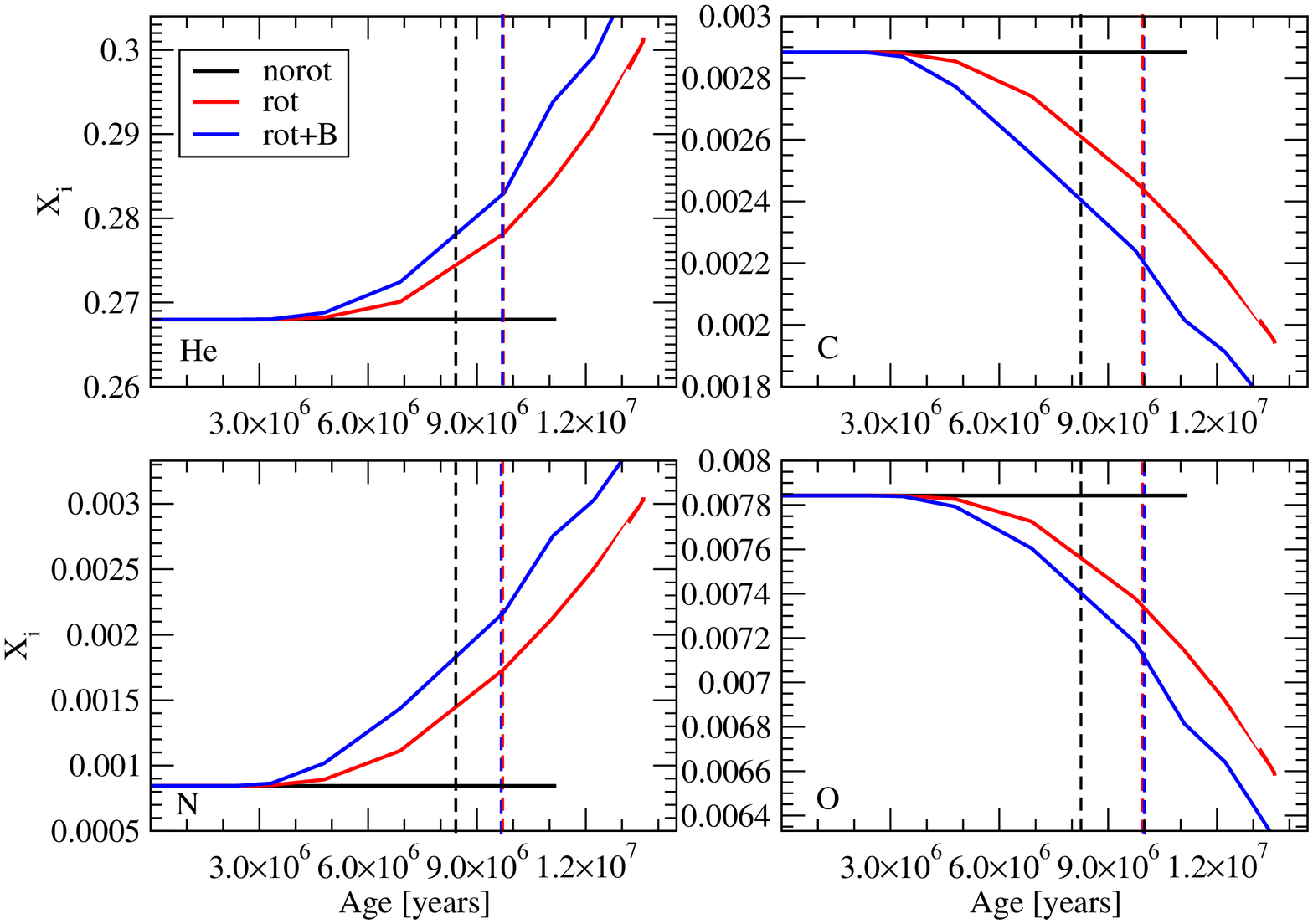}
\renewcommand\thefigure{A4} 
\caption{Evolution of the He (upper left panel), C (upper right panel), N (lower left panel) and O (lower right panel) surface mass fractions 
for a 15~$M_{\odot}$, $Z =$~0.014 in the case of zero rotation (black curves), initial rotation of 200~km~s$^{-1}$ (red curves) and initial
rotation of 200~km~s$^{-1}$ plus magnetic fields (blue curves) as calculated in MESA. 
The dashed vertical lines of the same colors represent the time when
the central hydrogen mass fraction was $\sim$~35\%, in accordance with Heger, Woosley \& Spruit (2005). The effects of the ST mechanism
are included for both the angular momentum transport and the chemical mixing for the rotating plus magnetic models.}
\end{center}
\end{figure}

\begin{figure}
\begin{center}
\includegraphics[angle=-90,width=18cm]{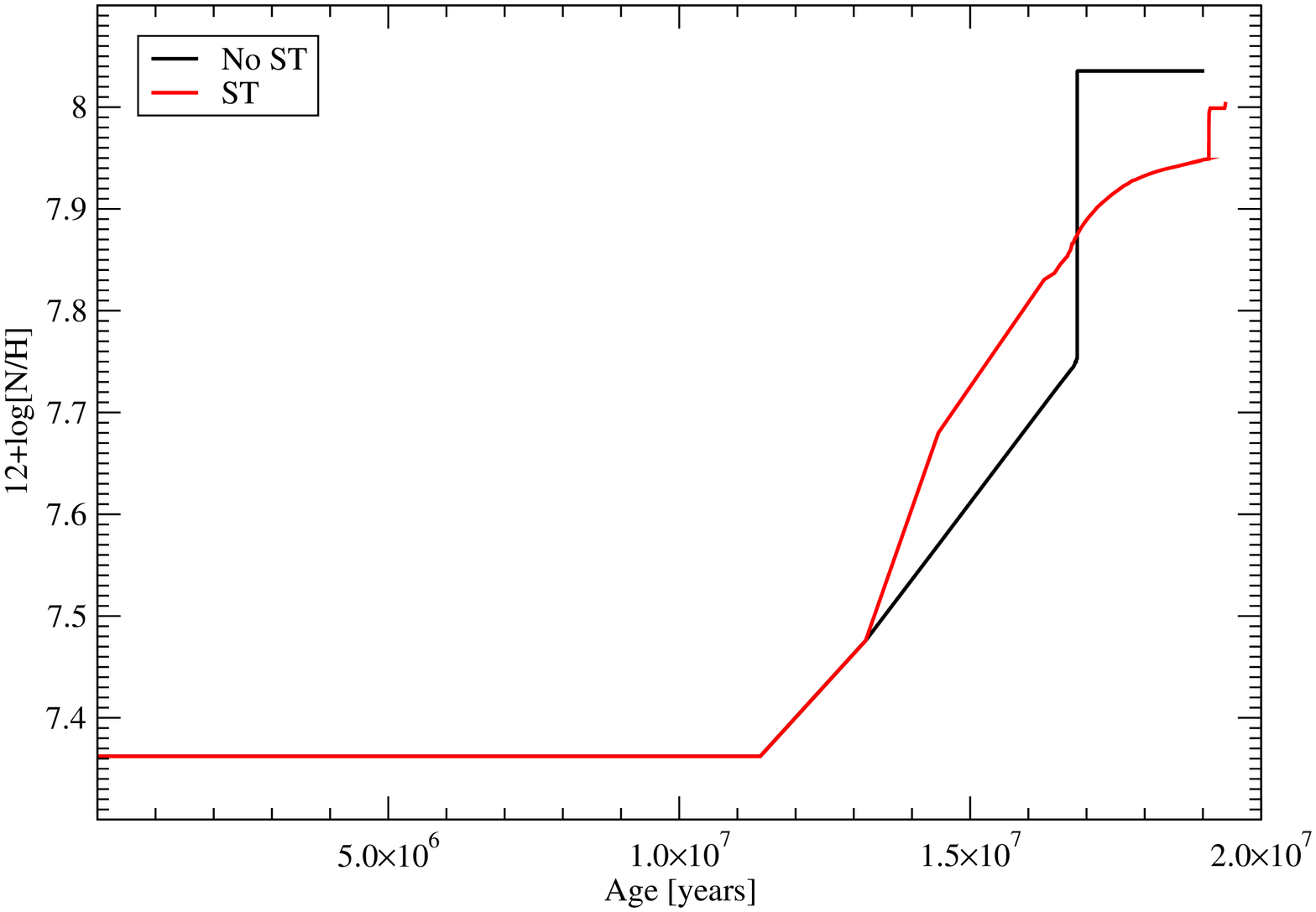}
\renewcommand\thefigure{A5} 
\caption{Evolution of surface nitrogen abundace with time for a 12~$M_{\odot}$, $Z =$~0.004 model rotating at 354.14~km~s$^{-1}$ for 
a case where the effects of the ST mechanism on chemical mixing are ignored (black curve) 
and for a case where those effects are included (red curve) as calculated by MESA. 
The effects of the ST mechanism on angular momentum transport were included for both models.}
\end{center}
\end{figure}

\begin{figure}
\begin{center}
\includegraphics[angle=-90,width=18cm]{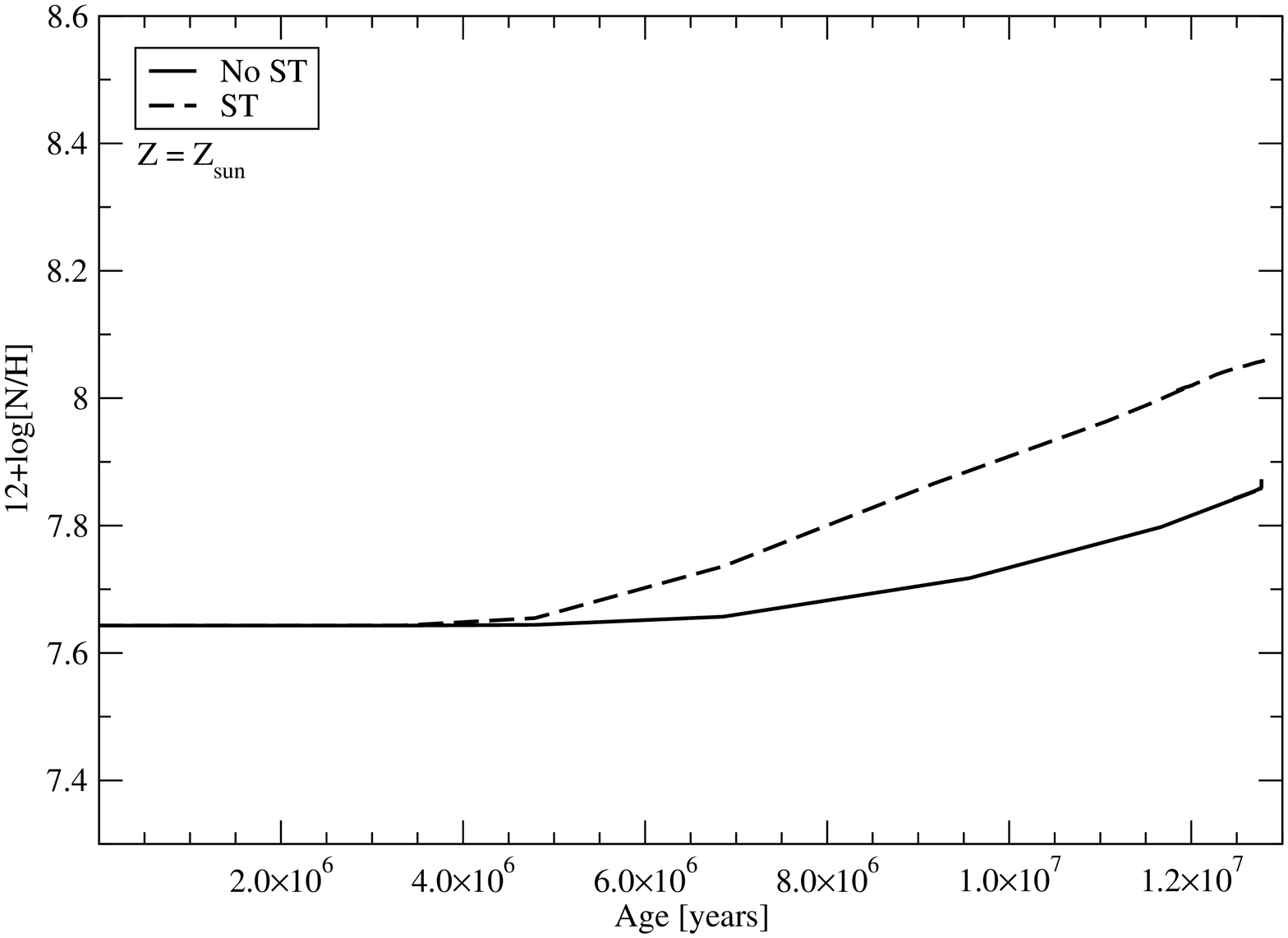}
\renewcommand\thefigure{A6} 
\caption{Evolution of 12+$\log[N/H]$ for a 15~$M_{\odot}$, $Z =$0.014 model rotating at 150~km~s$^{-1}$ in the case where
the effects of the ST mechanism on chemical mixing are ignored (solid black curve) 
and for a case where those effects are included (dashed black curve) as calculated by MESA.
The effects of the ST mechanism on angular momentum transport were included for both models.}
\end{center}
\end{figure}

\begin{figure}
\begin{center}
\includegraphics[angle=-90,width=18cm]{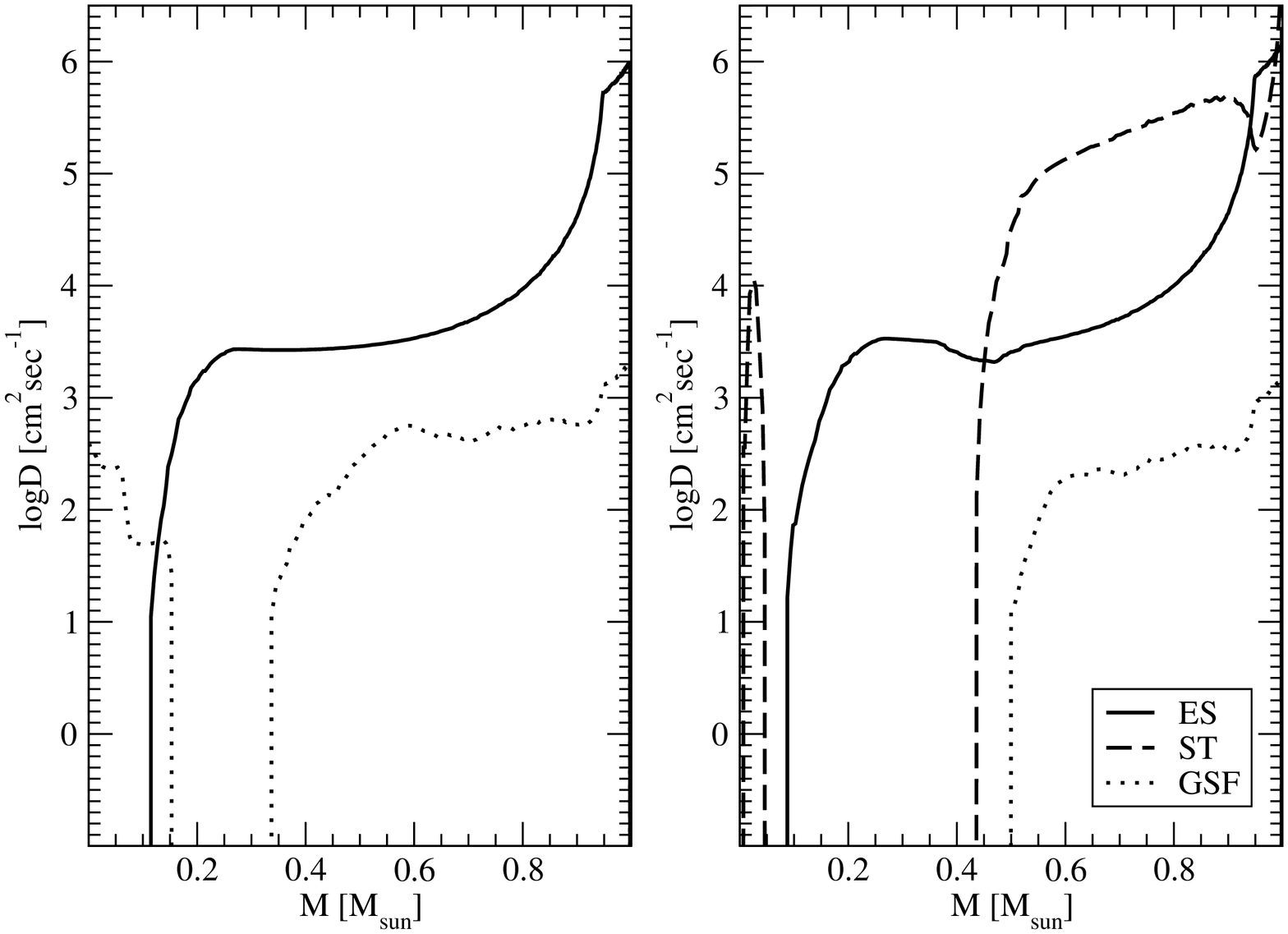}
\renewcommand\thefigure{A7} 
\caption{Distribution of the mixing coefficients due to the Eddington-Sweet circulation (ES; solid curves), 
the Goldreich-Schubert-Fricke instability (GSF; dotted curves) and the ST mechanism (dashed curves) for the 1~$M_{\odot}$ model
rotating at 30\% the critical value in a case where ST mixing is ignored (left panel) and in a case where it is included (right panel).
The effects of the ST mechanism on angular momentum transport were included in all cases.}
\end{center}
\end{figure}


\clearpage
\setcounter{table}{0}
\begin{deluxetable}{llllllllllllcccccc}
\tabletypesize{\tiny}
\tablewidth{0pt}
\tablecaption{Chemical properties of the models for low-mass binary secondaries studied in this work.}
\tablehead{
\colhead {$M_{\rm ZAMS}$~($M_{\odot}$)} &
\colhead {$\Omega/\Omega_{\rm crit,ZAMS}$} &
\colhead {$X_{\rm H,s,\tau_{\rm MS}}/X_{\rm H,s,i}$~$^{a}$} &
\colhead {$X_{\rm He,s,\tau_{\rm MS}}/X_{\rm He,s,i}$~$^{a}$} &
\colhead {$X_{\rm C,s,\tau_{\rm MS}}/X_{\rm C,s,i}$~$^{a}$} &
\colhead {$X_{\rm N,s,\tau_{\rm MS}}/X_{\rm N,s,i}$~$^{a}$} &
\\
}
\startdata
\hline
~&~&~&$Z =$~$Z_{\odot}$ &~&~\\
\hline
0.8  & 0.0  & 1.0000 &  1.0000 &  1.0000 &  1.0000  \\
0.8  & 0.3  & 0.9640 &  1.0881 &  0.8266 &  1.6979  \\
\hline
1.0   & 0.0 & 1.0000 &  1.0000 &  1.0000 &  1.0000  \\ 
1.0   & 0.3 & 0.8986 &  1.1535 &  0.5018 &  1.5072  \\
\hline
1.2   & 0.0 &  1.0000 &  1.0000 &  0.8202 &  1.0000  \\
1.2   & 0.3 &  0.9365 &  1.1520 &  0.5019 &  3.5157  \\
\hline
1.5   & 0.0 &  1.0000 &  1.0000 &  0.8202 &  1.0000  \\
1.5   & 0.3 &  0.9698 &  1.0723 &  0.5255 &  3.3124  \\
\hline
1.8  & 0.0 &  1.0000 &  1.0000 &  0.8202 &  1.0000  \\
1.8  & 0.3 &  0.9827 &  1.0414 &  0.5574 &  3.1625  \\
\hline  
~&~&~&$Z =$~0.1~$Z_{\odot}$ &~&~\\
\hline
0.8  & 0.0 & 1.0000  &  1.0000 &  1.0000 &  1.0000  \\ 
0.8  & 0.3 & 0.9158  &  1.2592 &  0.5754 &  2.6936  \\
\hline
1.0  & 0.0 & 1.0000  &  1.0000 &  1.0000 &  1.0000  \\ 
1.0  & 0.3 & 0.9153  &  1.2653 &  0.2915 &  3.7102  \\
\hline
1.2  & 0.0 & 1.0000  &  1.0000 &  1.0000 &  1.0000  \\
1.2  & 0.3 & 0.9101  &  1.2534 &  0.1943 &  4.1908  \\
\hline
1.5   & 0.0 &  1.0000 &  1.0000 &  1.0000 &  1.0000  \\
1.5   & 0.3 &  0.9610 &  1.1170 &  0.3182 &  3.7951  \\
\hline
1.8  & 0.0 &  1.0000 &  1.0000 &  1.0000 &  1.0000  \\
1.8  & 0.3 &  0.9516 &  1.0995 &  0.3618 &  3.2097  \\
\hline  
~&~&~&$Z =$~0.01~$Z_{\odot}$ &~&~\\
\hline
0.8  & 0.0 & 1.0000  &  1.0000 &  1.0000 &  1.0000  \\
0.8  & 0.3 & 0.8917  &  1.3472 &  0.4502 &  3.2118  \\
\hline
1.0  & 0.0 & 1.0000  &  1.0000 &  1.0000 &  1.0000  \\ 
1.0  & 0.3 & 0.8363  &  1.4405 &  0.0825 &  5.7059  \\
\hline
1.2  & 0.0 & 1.0000  &  1.0000 &  1.0000    &  1.0000\\
1.2  & 0.3 & 0.8860  &  1.3642 &  0.0687    &  5.2941\\
\hline
1.5   & 0.0 &  1.0000 &  1.0000 &  1.0000 &  1.0000  \\
1.5   & 0.3 &  0.8976 &  1.3049 &  0.1203 &  5.4235  \\
\hline
1.8  & 0.0 &  1.0000 &  1.0000 &  1.0000 &  1.0000  \\
1.8  & 0.3 &  0.9254 &  1.2287 &  0.1443 &  5.9529  \\
\enddata 
\tablecomments{$^{a}$ The ratio of the ZAMS surface mass fraction of a specific element to the surface mass fraction of the same element
at the end of the MS.}
\end{deluxetable}

\end{document}